\documentclass[fleqn,usenatbib]{mnras}

\usepackage{newtxtext,newtxmath}

\usepackage[T1]{fontenc}

\DeclareRobustCommand{\VAN}[3]{#2}
\let\VANthebibliography\thebibliography
\def\thebibliography{\DeclareRobustCommand{\VAN}[3]{##3}\VANthebibliography}

\usepackage{graphicx}	
\usepackage{amsmath}	
\usepackage{booktabs}
\usepackage[table,xcdraw]{xcolor}
\usepackage{subcaption}
\usepackage{stfloats}

\newcommand{\hi}{H\,{\sc{i}}\,}

\newcommand{\msol}{M$_{\odot}$}
\newcommand{\kms}{km\,s$^{-1}$}

\title{Decoding the star forming properties of gas-rich galaxy pairs}

\author[Bok et al.]{
J. Bok,$^{1,2}$\thanks{E-mail: jamie@ast.uct.ac.za}
M.E. Cluver,$^{3,4}$
T.H. Jarrett, $^{2}$
R.E. Skelton,$^{1}$
M.G. Jones, $^{5}$ 
\newauthor
 \hspace{1mm}and L. Verdes-Montenegro $^{5}$
\\
$^{1}$South African Astronomical Observatory, Observatory, Cape Town, 7935,  South Africa \\
$^{2}$University of Cape Town, Rondebosch, Cape Town, 7700, South Africa \\
$^{3}$Centre for Astrophysics and Supercomputing, Swinburne University of Technology, Hawthorn, Victoria 3122, Australia \\
$^{4}$Department of Physics and Astronomy, University of the Western Cape,Robert Sobukwe Road, Bellville, South Africa\\
$^{5}$Instituto de Astrof\'isica de Andaluc\'{i}a (IAA-CSIC), Glorieta de la Astronom\'{i}a, 18008 Granada, Spain}

\date{Accepted XXX. Received YYY; in original form ZZZ}

\pubyear{2015}

\begin{document}
\label{firstpage}
\pagerange{\pageref{firstpage}--\pageref{lastpage}}
\maketitle

\begin{abstract}
We extend the analysis of Bok et al. (2020) in which the H{\textsc i~}content of isolated galaxies from the AMIGA sample and selected paired galaxies from ALFALFA were examined as a potential driver of galaxy location on the \textit{WISE} mid-infrared SFR-M$\star$ sequence. By further characterizing the isolated and pair galaxy samples, i.e. in terms of optical galaxy morphology, a more detailed and quantitative description of local galaxy environment by way of the local number density ($\eta$) and tidal strength (Q) parameters, star formation efficiency (SFE$_{\textrm{H{\textsc i}}}$), and H{\textsc i~}integrated profile asymmetries, we present plausible pathways for the broadening of the pair sample H{\textsc i~}deficiency distribution towards both high and low deficiencies compared to the narrower isolated galaxy sample distribution (i.e. $\sigma_{\rm{PAIRS}} = 0.34$ versus $\sigma_{\rm{AMIGA}} = 0.28$). We associate the gas-rich tail of the pair deficiency distribution with the highest Q values, large profile asymmetries, and low SFEs. From this we infer that merger activity is enhancing gas supplies, as well as disrupting the efficiency of star formation, via strong gravitational torques. The gas-poor wing of the deficiency distribution appears to be populated with galaxies in denser environments (with larger $\eta$ values on average), more akin to groups. Despite our gas-rich selection criterion, there is a small population of early-type galaxies in the pair sample, which primarily fall in the positive deficiency wing of the distribution. These results suggest that a combination of a denser galaxy environment, early-type morphology, and higher stellar mass, is contributing to the broadening of the deficiency distribution towards larger deficiencies. 

\end{abstract}




\section{Introduction}

With SKA pathfinders and precursors now operational, and H{\textsc i~}(atomic hydrogen) science poised on the edge of a vast new discovery space, studies of the H{\textsc i~}properties of galaxies have become increasingly accessible, and our understanding of the important role H{\textsc i~}content plays in galaxy evolution greatly advanced. As the fuel for star formation, H{\textsc i~}is now understood to be a key piece of the galaxy bimodality puzzle; the transition from a star-forming spiral to a passive elliptical, i.e. the ultimate cessation of star formation, is naturally linked not only to the availability of star fuel, but also to how efficiently a galaxy can convert its fuel to stars. A useful diagnostic for examining the processes driving star formation (SF) in galaxies is the SFR-M$_{{\rm{star}}}$ sequence, otherwise referred to as the galaxy main sequence (GMS), or star-forming main sequence (SFMS). The relation shows a general trend of star-formation rate (SFR) increasing with stellar mass \citep{Noeske2007, Bouche2010, Lee2015}, but not without significant scatter \citep{Hall2018}. Galaxies are observed to jump up off the SFMS as they undergo triggered bursts of rapid SF (starburst events), and drop off the SFMS to due quenching events. Recent studies have shown H{\textsc i~}to be an important driver of galaxy location on the SFMS \citep{Saintonge2016,Bok2020}, with bursts of enhanced SF associated with an increased availability of H{\textsc i~}\citep{Ellison2010, Moreno2020}, and low SF galaxies deficient in H{\textsc i~}\citep{Bok2020}. Galaxy environment is also implicated in many studies as linked causally to both the elevation and suppression of SF \citep{Ellison2010, Xu2010,Cochrane2018,Pearson2019, Moon2019, Moreno2020}. Our current understanding is that environment drives SF via its impact on the H{\textsc i~}content of galaxies. The fragile nature of the diffuse, extended, and high angular momentum H{\textsc i~}disk component of a galaxy makes it particularly sensitive to the local environment in which it resides, and the significant impact galaxy environment has on both H{\textsc i~}content and distribution is well documented \citep{Jones2018,Cortese2021}.\\\\
\noindent Numerous studies have revealed galaxies in the centers of clusters to be gas-poor compared to less densely populated environments \citep{Chamaraux1980,Giovanelli1985,Solanes2001,Chung2009,Denes2014}, with ram-pressure stripping generally attributed as the dominant gas removal mechanism in such environments \citep{GunnGott1972}. \cite{Chung2009} conducted an H{\textsc i~}morphology census of galaxies in the Virgo cluster at different radii from the center and found that close to the core (within $~0.5$ Mpc), the H{\textsc i~}disks of the galaxies were generally smaller than their optical counterparts, and in many cases these galaxies were also shown to have displaced gas--- potentially evidence of current stripping, or the in-fall of previously stripped gas. Conversely, gas rich galaxies with gas disks extending well beyond their stellar components were found to dominate in the cluster outskirts. At intermediate distances from the cluster core a significant number of galaxies had long one-sided H{\textsc i~}tails pointing away from the center, and the authors proposed that these galaxies were in the process of falling into the cluster core for the first time.\\\\
Galaxies in groups have also been reported as being H{\textsc i~}deficient compared to field galaxies \citep{Verdes-Montengero2001,Kilborn2009,HessWilcots2013}, with the cores of groups becoming increasingly H{\textsc i~}deficient as group membership increases \citep{HessWilcots2013}. \cite{HessWilcots2013} present compelling evidence that galaxies transition from gas-rich to gas-poor in the group environment, via either star formation, tidal interactions, starvation, viscous stripping, or ram pressure stripping. Interestingly, X-ray studies of the IGM (intragroup medium) in Compact Groups \citep{Hickson1982} suggest that ram-pressure stripping is not the dominant gas removal mechanism in these systems \citep{Rasmussen2008}. High quality single-dish Green Bank Telescope (GBT) observations of the groups revealed a diffuse H{\textsc i~}component suggestive of tidal stripping \citep{Borthakur2010,Borthakur2015A}, and it is generally suggested that H{\textsc i~}deficiency in groups is primarily a consequence of tidal stripping via galaxy-galaxy or galaxy-group interactions \citep{Kilborn2009}, although a recent review by \cite{Cortese2021} suggests full quenching is the combined result of various gas-removal mechanisms, acting simultaneously, or at different stages throughout the galaxy life cycle. Observing the diffuse H{\textsc i~}(intra-group clouds, tidal tails, swept back disks), that would allow us to concretely differentiate between the various gas-removal mechanisms, is currently only feasible for a few nearby groups due to the costly observing requirements (high resolution, time consuming). SKA1-MID (the first stage of SKA) will significantly outperform current facilities. For the same integration time and angular resolution, SKA1-MID will reach column density limits 3 times deeper than what is currently possible, probing IGM column densities for the first time, and resolving individual H{\textsc i~}clouds in the Local Volume \citep{Popping2015}. \\\\ 
Simulations of merging galaxy pairs demonstrate the pair environment to have a significant impact on the H{\textsc i~}content of galaxies, but different from the cluster and group environment. Numerical simulations show the gravitational interplay of merging galaxy pair members to generate internal asymmetries and instabilities that funnel gas inwards, producing the well known central concentration \citep{Mihos1996}, and more recently, \cite{Hani2018} used a zoom-in simulation to show that a major merger can also re-distribute material into the circumgalactic medium (CGM), and increase total hydrogen covering factors by a factor of 1-1.25 for up to $\sim3$ Gyr. Parsec-scale galaxy major-merger simulations implemented by \cite{Moreno2020} demonstrated that close galaxy encounters can elevate both H{\textsc i~}and H$_2$ gas reserves, and thereby enhance SF in both pair members. Alongside simulations, a re-distribution of gas as tidal tails and central molecular gas concentrations is routinely observed in merging galaxies \citep{Hibbard1999,Tacconi1999,Koribalski2004,Yamashita2017}, and \cite{Ellison2018} find merging galaxies to be on average gas-rich compared to control samples. \\\\
In order to truly disentangle the impact of external influence from so called secular, i.e. steady state, evolution, we need a non-interacting control sample. To this end we defer to the AMIGA project (Analysis of the interstellar Medium in Isolated GAlaxies; \cite{VerdesMontenegro2005}), who provide the most comprehensive multi-wavelength study of a well-defined sample of isolated galaxies, and therefore the best possible estimate of secular evolution to date. In our previous paper \citep{Bok2020} we contrasted the H{\textsc i~}deficiency (and gas fraction) distribution of ALFALFA galaxies in close pairs with isolated galaxies from the AMIGA sample, and observed a broadening of the pair sample H{\textsc i~}deficiency (/gas fraction) distribution in both directions, towards higher and lower deficiencies (/gas fractions), relative to the isolated galaxy sample. The pair finding method used in \cite{Bok2020} was a first order definition of the environment since it required only that a target galaxy have at least one close companion, triples and groups are not excluded by the criteria. Visual inspection of the pair sample indicated that the pairs are located in environments of varying density. Since we know that different environments impact the H{\textsc i~}content and distribution in galaxies differently, we propose that the broadening of the pair H{\textsc i~}deficiency distribution towards both higher and lower H{\textsc i~}deficiencies may potentially be explained by the presence of a variety of local environments. \\\\
For these reasons we have gone a step further and characterised in finer detail the local environment of the pair sample by invoking the use of \cite{Verley2007a}'s local number density parameter, $\eta$, and tidal influence estimator, Q, which provide complementary information about the local environment in the vicinity of target galaxies. $\eta$, or $\eta_k$, probes the local vicinity of a target galaxy by taking into account the distance to its $kth$ nearest neighbour, and is defined as: $$\eta_k\propto log\frac{k-1}{V(r_k)}$$ where $r_k$ is the distance to the $kth$ nearest neighbour measured in arcminutes. To probe the local vicinity of the target galaxy, that is, the vicinity in which principal perturbers can be expected to lie, $k$ is set to equal 5, or less should there be too few neighbours in the field. To mitigate contamination by background galaxies, only neighbours of similar size are considered in the calculation (i.e. neighbours with diameters between 0.25 and 4 times the target galaxy)\citep{Verley2007a}. Using formalism developed by \cite{Dahari1984}, the tidal strength parameter, Q, then estimates the strength of the tidal forces exerted on the target galaxy by its k neighbours, and is proportional $M_iR_{ip}^{-3}$, where $M_i$ is the mass of the $ith$ neighbour and $R_{ip}$ is its distance from the primary. Q is ultimately defined as the logarithm of the ratio of the external tidal force to internal gravitation binding energy, i.e.: $$Q=log(\sum_{i}Q_{ip})$$ where $$Q_{ip}\equiv\frac{F_{tidal}}{F_{bind}}\propto (\frac{M_i}{M_p})(\frac{D_p}{S_ip})^3$$ where $M_{i,p}$ is the mass of the $ith$ neighbour and the primary respectively, $D_p$ is the diameter of the primary, and $S_{ip}$ is the projected separation \citep{Verley2007a}. For both the calculation of $\eta$ and Q we draw neighbours from the SDSS photometric catalogue, and make use of photometric r-band magnitudes to serve as a proxy for stellar mass. 
Not only can we gauge the number density of local companions, but by taking into account their stellar masses via the Q parameter we are also able to assess how susceptible the target galaxy is to the external gravitational influence of said candidate perturbers. 
We also use the SDSS galaxy group catalogue of \cite{Lim2017} to search for existing groups associated with the pair sample. Furthermore, we conducted a visual classification of the pair member optical morphologies to investigate how morphology might be correlated to the observed differences in H{\textsc i~}content we measured between our paired and isolated galaxy samples on the SFMS, and we use star-formation efficiencies to probe the mechanisms responsible for driving galaxies off the SFMS towards low SFRs, despite having relatively large gas reservoirs. Finally, we compute the H{\textsc i~}profile asymmetries of our pair sample and examine them as a function of local environment ($\eta$, Q, and number of group members for the \cite{Lim2017} groups), to test the popular proposition that the degree of profile asymmetry and density of environment positively correlate \citep{Espada2011,Scott2018,Bok2019}. \\ This paper is organised as follows: In Section 2 we describe and characterise our isolated and pair samples in terms of the $\eta$ and Q environment parameters (\S2.1), as well as group membership for the pair sample in Section 2.1.1. We present morphologies for the pair and isolated samples in Sections 2.2 and 2.3 respectively. In Section 3 we examine \hi deficiency as a function of both $\eta$ and Q (\S3.1), and galaxy morphology (\S3.2). In Section 3.3 we present and investigate the role of star formation efficiencies in driving trends on the SFMS, and their relationship to both \hi deficiency and gas fraction. In Section 3.4 we present and compare our pair profile asymmetries with the literature, and investigate their relationship with $\eta$ and Q, group membership, and stellar mass. We discuss our results in context of the literature in Section 4, and summarise our main results and conclusions in Section 5. Throughout this paper we adopt a $\Lambda CDM$ cosmology with $H_0=70 $ \kms Mpc$^{-1}$, $\Omega_{M}= 0.3$, and $\Omega_{\Lambda} = 0.7$.

\section{Sample: isolated and paired galaxies}
This paper uses the pair and isolated galaxy samples defined in \cite{Bok2020}, where full details of the selection criteria can be found in \S 3.2. We include here a summary of the relevant details. The close pair galaxy sample admits H{\textsc i-}rich (M$_{\textrm{H\textsc i}}$>10$^9$ M$_{\odot}$) galaxies from the $\alpha70$ ALFALFA catalogue with at least one similarly gas-rich companion within 100 kpc and 1000\kms (as per the \cite{Robotham2014} close pair criteria). As our sample is drawn from a blind H{\textsc i~}survey we do not expect it to be representative of the global galaxy population, but rather biased towards gas-rich, star-forming systems, most notably in the low stellar mass regime \citep{Saintonge2016}. Our isolated galaxy sample comprises a selection of gas-rich (M$_{{\rm{H\textsc i}}}>10^9$\msol) galaxies from the complete AMIGA H{\textsc i~}science sample that have been flagged as reliably isolated, and with high quality H{\textsc i~}profiles (544 galaxies in total). By admitting only gas-rich galaxies into our isolated and pair samples we preferentially select star-forming late-type galaxies. To mitigate the effect of missing low stellar mass galaxies in the AMIGA sample we implement a (low) stellar mass cut of M$_{\star}>10^{8.5}$ \msol \hspace{1mm} across both the isolated and H{\textsc i~}pair samples, which allows for a more fair comparison between the two samples. This lower limit on stellar mass additionally acts to reduce the effects of bias in the pair sample towards gas-rich systems in the low stellar mass regime. Our final isolated galaxy sample comprises 481 galaxies, while 531 galaxy pair members in our H{\textsc i~}pair sample survive the stellar mass cut. A summary of the sample criteria are listed in Tables 3.1 and 3.2 in \S 3.2 of \cite{Bok2020}.

\subsection{Quantification of the broader `pair' environment: eta and Q}

\cite{Verley2007a} quantified the degree of isolation of \cite{Karachentseva1973}'s CIG galaxies using a local number density of neighbouring galaxies ($\eta$), and the tidal strength exerted by these neighbouring galaxies on the CIG galaxy (Q). By taking into account both the number of neighbours and their masses, $\eta$ and $Q$ work together to provide a comprehensive description of the local environment in the vicinity of each CIG galaxy, well suited to assessing isolation. Galaxies that are truly isolated from external influence have low $\eta$ and $Q$ values. Conversely, when $\eta$ and $Q$ are both high, we know that the evolution of the target galaxy can be, or already has been, impacted by the environment. The latter is therefore not representative of evolution in isolation. $\eta$ and $Q$ can also be used to differentiate between environments in which a single close neighbour (low $\eta$) has a considerable tidal influence (high  $Q$) on a target galaxy, and environments densely populated (high $\eta$) with relatively small (low mass) galaxies (low Q). \cite{Verley2007a} also demonstrated $\eta$ and $Q$ to be effective at distinguishing isolated galaxies from galaxies in denser environments by comparing them with triplets from the Karachentseva's catalogue \citep{Karachentseva1973}, compact groups from the Hickson catalogue (HCG; \cite{Hickson1982}) and Abell clusters (ACO; \cite{Abell1958}). \\

\begin{figure*}
    \centering
    \includegraphics[scale = 0.55]{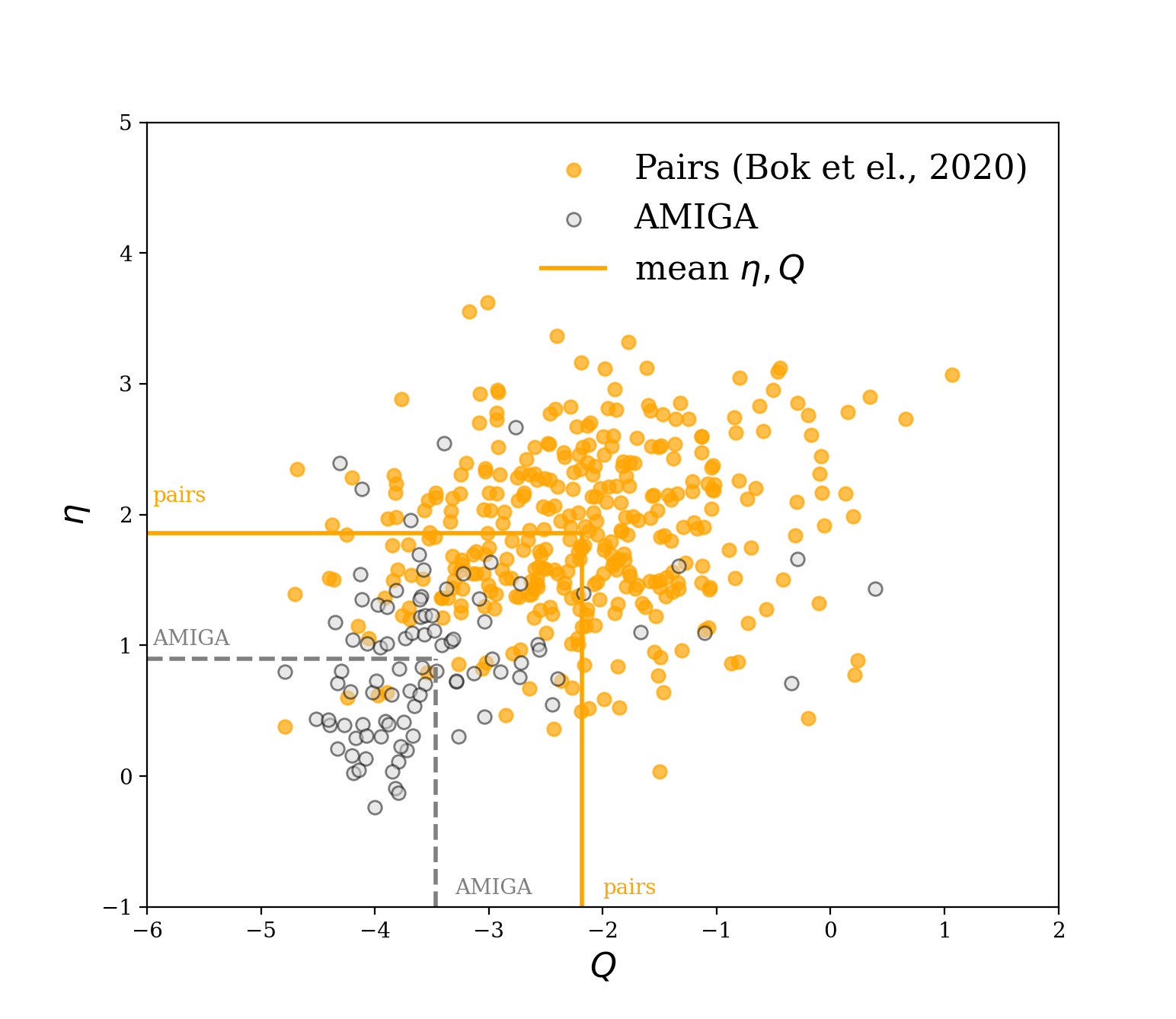}
    \caption{H{\textsc i~}pair environment: local density ($\eta$) and tidal strength (Q) of the pairs (orange) and the isolated field galaxies (grey). The solid orange line marks the location of the mean pair sample $\eta$ and $Q$ values. A clear separation of the pair and field galaxies is discerned in this diagram.} 
    \label{fig:1}
\end{figure*}

\noindent In Figure \ref{fig:1} we present the $\eta$ and $Q$ measurements for our pair sample (orange points) and contrast them with our isolated sub-sample of AMIGA galaxies (gray points) on the $\eta-Q$ plane. Mean values for each parameter are shown as orange and gray dashed lines for the respective samples. We note that the H{\textsc i~}pair sample is distinctly separate from isolated galaxies on the diagram (demonstrating the efficacy of the parameters to distinguish between the two environments). Furthermore, many of our pair galaxies likely reside in more densely populated environments according to the $\eta$ and $Q$ values of the \cite{Karachentseva1973} triplets, Hickson Compact Groups, and Abell Cluster galaxies shown in Figure 6 of \cite{Verley2007a}. These metrics, however, are not entirely equivalent when based on different photometric surveys, so we cannot compare directly.  

\subsubsection{Groups in the pair sample}
The possibility that some of our pair sample galaxies may additionally be embedded in groups was noted by \cite{Bok2020}, and is also suggested by the broad range of environments spanned by the pair sample on the $\eta-Q$ plane (Figure \ref{fig:1}). \cite{Lim2017} produced a galaxy group catalogue of the low-redshift Universe using a compilation of large redshift surveys, and cross-matching with their SDSS group catalogue confirms that 85 of our pair galaxies are located in groups, with at least N>2 members. The halo-based group finding methods employed by \cite{Lim2017} are based on those of \cite{Yang2007}, however with an improved halo mass assignment, which extends the group catalogue to include more low-mass systems (see \cite{Lim2017} for full details). It is important to note that the \cite{Lim2017} groups do not necessarily identify all the pairs in our sample that are in groups.

\subsection{Pair member morphologies}
\begin{figure}
    \centering
    \includegraphics[scale = 0.5]{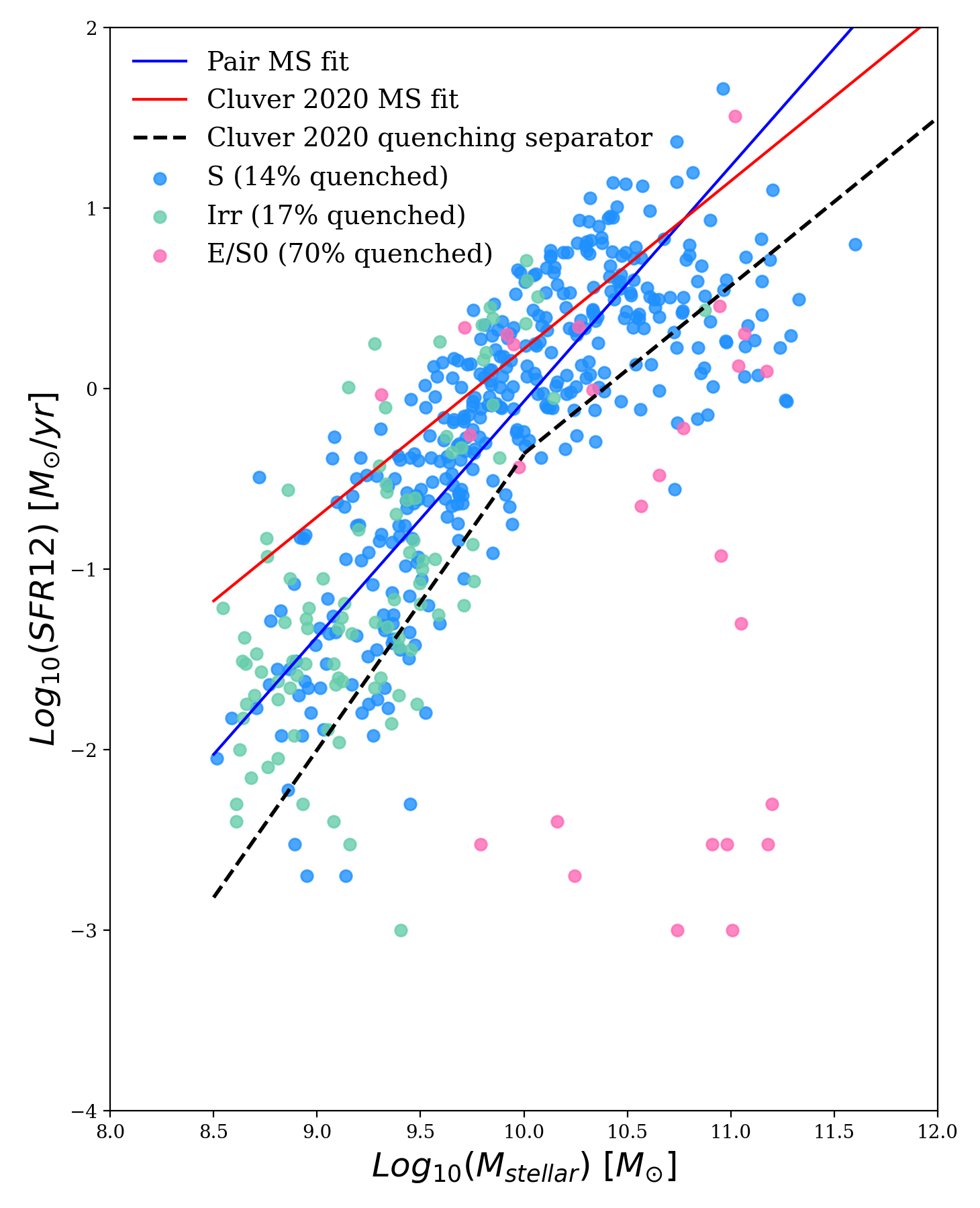}
    \caption{Galaxy morphology on the pair SFMS. Here blue and green represent the spiral and irregular populations respectively, i.e. the late-type population, and pink the spheroidal/elliptical and S0 (early-type) population. The fraction of quenched galaxies in each morphology group, determined as per the quenching separator from \protect\cite{Cluver2020} (black dashed line), is indicated as a percentage next to each category in the legend. The SFMS fit from \protect\cite{Cluver2020} for the non-grouped GAMA sample is shown for reference in red.}
    \label{fig:morphMS}
\end{figure}

The link between SF activity and galaxy colour is well-established, with the majority of typical galaxies generally falling into two categories: the "red sequence" and the "blue cloud" \citep{Strateva2001,Kauffmann2003,Baldry2004,Balogh2004}. Hubble Type (galaxy morphology) roughly correlates with these two populations. "Red sequence" galaxies are generally high mass, passive, and early-type (bulge-dominated) in morphology, whereas galaxies in the "blue cloud" are typically less massive, active (star forming), and of late-type (disk dominated) morphology. To assess the role of morphology as a driver of our results, we conducted a visual classification of each galaxy using optical SDSS DR12 images, which were available for 497/524 of our pair member galaxies. Of these, each galaxy was classified by JB, MC and TJ, as either early-type (typically spheroids/ellipticals and S0s) or late-type (spirals and irregulars). In most cases ($\sim85\%$ of the time) the classifications were in agreement. Disagreements were appropriately reclassified by vote. \\\\In Figure \ref{fig:morphMS} we plot the pair SFMS colour-coded by morphology and see the familiar trend of late-type galaxies dominating the SF regions of the plot, namely on and around the SFMS (blue points), whereas as the small population of early-type galaxies are mostly high mass with very low SFRs (pink points). Our visual morphology classifications, which show the pair population to be late-type (spiral) galaxy dominated, are in good agreement with the MIR morphology diagnostics presented in \cite{Bok2020}, namely the {\it{WISE}} infrared colour-colour diagram (Figure 4 of \cite{Bok2020}), which delineates 4 general galaxy types, and {\it{WISE}} B/T measurements (Figure 14 of \cite{Bok2020}), which is sensitive to the bulge and disk-dominated systems. \\\\ Using the \cite{Cluver2020} quenching separator derived using ungrouped galaxies in the GAMA survey (dashed black line in Figure \ref{fig:morphMS}) to define the fraction of quenched galaxies in each population, we find a strong correlation between our visual morphology classification scheme and SF activity. Our early-type sample is largely passive (70$\%$ quenched), and our late-type sample is predominantly SF (14$\%$ quenched). Our visual inspection yielded a total of 377 spiral (76 $\%$), 95 irregular (19$\%$), and 25 early-types (5$\%$) galaxies. \cite{Kelvin2014} find the local Universe to be approximately 70$\%$ disk dominated, and 30$\%$ spheroid dominated based on 3727 galaxies from the Galaxy and Mass Assembly (GAMA) survey to z <0.08. The relative absence of spheroids in our pair sample compared to the GAMA sample is a consequence of our gas-rich requirement, which strongly selects active disk galaxies, and excludes passive/early-type galaxies. \\\\ While galaxy morphology is clearly linked to location on SFMS, as is to be expected, we note that by not controlling for it in the pair sample we have allowed for the admittance of a particularly interesting sub-sample of star-forming early-type galaxies on/close to the SFMS (pink points), amongst a population of spiral disk galaxies that appear to be rolling off the SFMS. These galaxies are very likely in the process of, or soon to be, transitioning onto the passive sequence, and therefore present a unique opportunity to capture galaxy evolution in action. It is also interesting to note that our gas-rich selection has not entirely precluded low star-forming early-types from our sample, which we would otherwise expect to be gas-poor. Going forward we therefore opt to not exclude our early-type sample from our analysis, but rather track them, and discuss their impact on any potential trends. 

\subsection{AMIGA morphologies for the isolated sample}
Morphologies for the full AMIGA sample are available in Table 1 of \cite{Fernandez2012}, who revised the classifications performed by \cite{Sulentic2006} using CCD images from either SDSS or their own. Of the AMIGA galaxies that form our sample of isolated galaies in \cite{Bok2020}, 466 (97 $\%$) are classified as spirals, 8 as irregular (2 $\%$), and 4 (1 $\%$) as elliptical/S0. Naturally, since early-type galaxies are typically found in the centers of galaxy clusters (the galaxy-morphology density relation), their fraction in the AMIGA sample is very low.

\section{Decoding the H{\textsc i~}deficiency distribution of paired galaxies: a closer look at environment and morphology}
  \subsection{H{\textsc i~}deficiency on the eta-Q plane} 
\begin{figure}
    \centering
    \includegraphics[scale = 0.53]{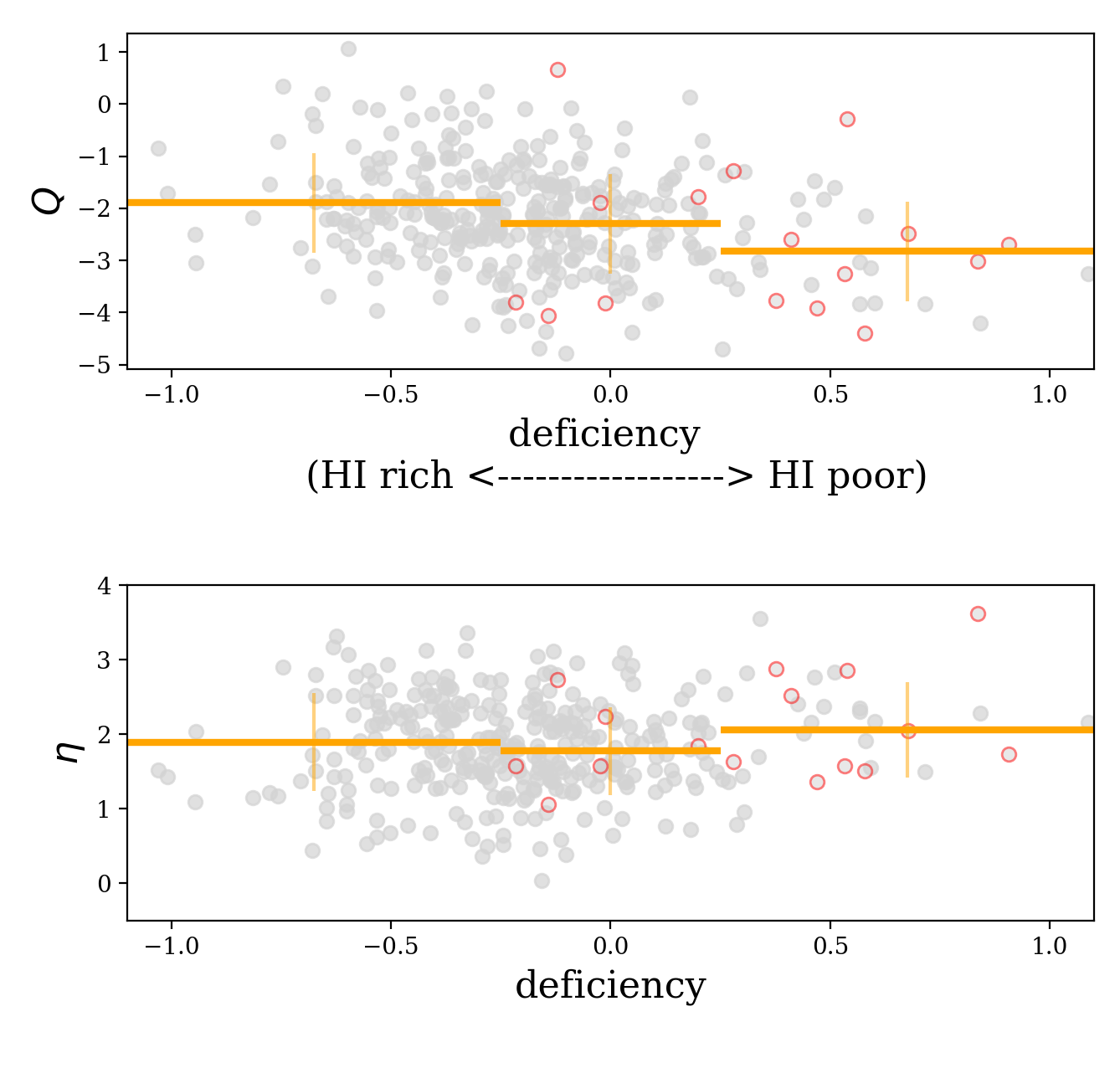}
\caption{Deficiency as a function of tidal strength ($Q$) (top) and $\eta$ (bottom) for the pair sample. Orange horizontal lines show the mean $Q$ and $\eta$ per deficiency bin, namely gas rich (DEF<-0.25), 'normal' \hi content (-0.25<=DEF<0.25), and \hi deficient (DEF >= 0.25)}, with the early-type galaxies circled in red.
\label{fig:etaq2panels}
\end{figure}

The H{\textsc i~}deficiency values presented here are those calculated by \cite{Bok2020}, which make use of an updated H{\textsc i-}M$\star$ scaling relation computed on a sub-sample of AMIGA galaxies with both H{\textsc i~}and \textit{WISE} data (See Equation 3 and Figure 5 in \cite{Bok2020}). We reiterate here how well suited the AMIGA sample is to providing a baseline for H{\textsc i~}`normalcy', and thus deficiency, due to the strict isolation criteria adhered to, and rigorous nature in which the sample has been studied by the AMIGA team- the AMIGA sample is arguably the most well-defined sample of isolated galaxies available in the literature, and thus the most appropriate control sample with which to gauge environmental impact on any galaxy property, including H{\textsc i~}content. The very purity of the AMIGA sample, that which makes it an ideal environmental control, however, comes at the price completeness in both stellar mass (missing low mass galaxies) and galaxy morphology (containing few early-types). While various quantities have been used to compute the H{\textsc i~}scaling relation (and hence deficiency), namely optical diameter \citep{Chamaraux1980,Haynes1984, Jones2018}, colour and structural properties \citep{Catinella2010}, and specific angular momentum \citep{Obreschkow2016}, the relationship between stellar and H{\textsc i~}mass is also commonly used (e.g. \cite{Catinella2010,Huang2012,Denes2014,Maddox2015, Parkash2018}). The updated AMIGA H{\textsc i-}M$\star$ scaling relation can thus be directly compared with many existing galaxy samples in the literature to directly assess the impact of galaxy environment.\\ \cite{Cortese2021} provide a comprehensive review of the various techniques for computing H{\textsc i~}deficiency, noting their individual strengths and limitations. They ultimately discourage making use of definitions based solely on stellar mass or luminosity due to their large scatter, which is strongly dependent on SFR, and because they often differ from the classical optical size methods. \cite{Jones2018}, however, found that the definitions for H{\textsc i~}deficiency were consistent when using either optical diameter ($D_{25}$) or luminosity ($L_B$), and that the scatter was the same. They attribute the disagreement between the \cite{Haynes1984} $D_{25}$ and $L_B$ relations to the fitting method used, i.e. the ordinary least squares method (OLS, as opposed to the the maximum likelihood estimator (MLE) used by \cite{Jones2018} and \cite{Bok2020}), which did not fully account for the (large) uncertainties in the luminosities, and thus produced a bias fit, which impacts the luminosity-based relation much more than the diameter-based one.\\ \cite{Cortese2021} also point out that optical size methods are not necessarily well suited to identify stripping events where gas and stars are simultaneously affected (such as tidal stripping). In this work we specifically study close, potentially interacting, \textit{gas-rich} pairs, as we are particularly interested in capturing such tidal events, which may well present as anomalous H{\textsc i~}content, and/or profile asymmetries. Close pairs are additionally usually star-forming and dusty by nature. By making use of mid-infrared derived stellar masses we circumvent the issue of dust obscuration that we would certainly encounter in the optical regime if we were to try and measure optical diameters. \\In Figure \ref{fig:etaq2panels} we show the H{\textsc i~}deficiency of the close pair members as a function of tidal strength, $Q$, (top) and number density, $\eta$, (bottom) separately. The mean values of $Q$ and $\eta$ per deficiency bin are shown as orange horizontal lines. This plot indicates a tentative trend of decreasing $Q$ with increasing deficiency, and no trend between $\eta$ and deficiency. These results suggest that tidal influence is potentially better suited to predicting H{\textsc i~}content than density of environment, and demonstrate that a galaxy living in a densely populated environment is not necessarily subject to a larger tidal pressure than a galaxy living in a less densely populated environment- the mass of neighbours is important. Since there is no positive correlation between Q and H{\textsc i~}deficiency, we rule out tidal forces as a dominant mechanism for gas-depletion in our pair sample. The small E/S0 population (circled in red on the plot), do not appear correlated with either $Q$ or $\eta$.   

\begin{figure}
    \centering
    \includegraphics[scale = 0.55]{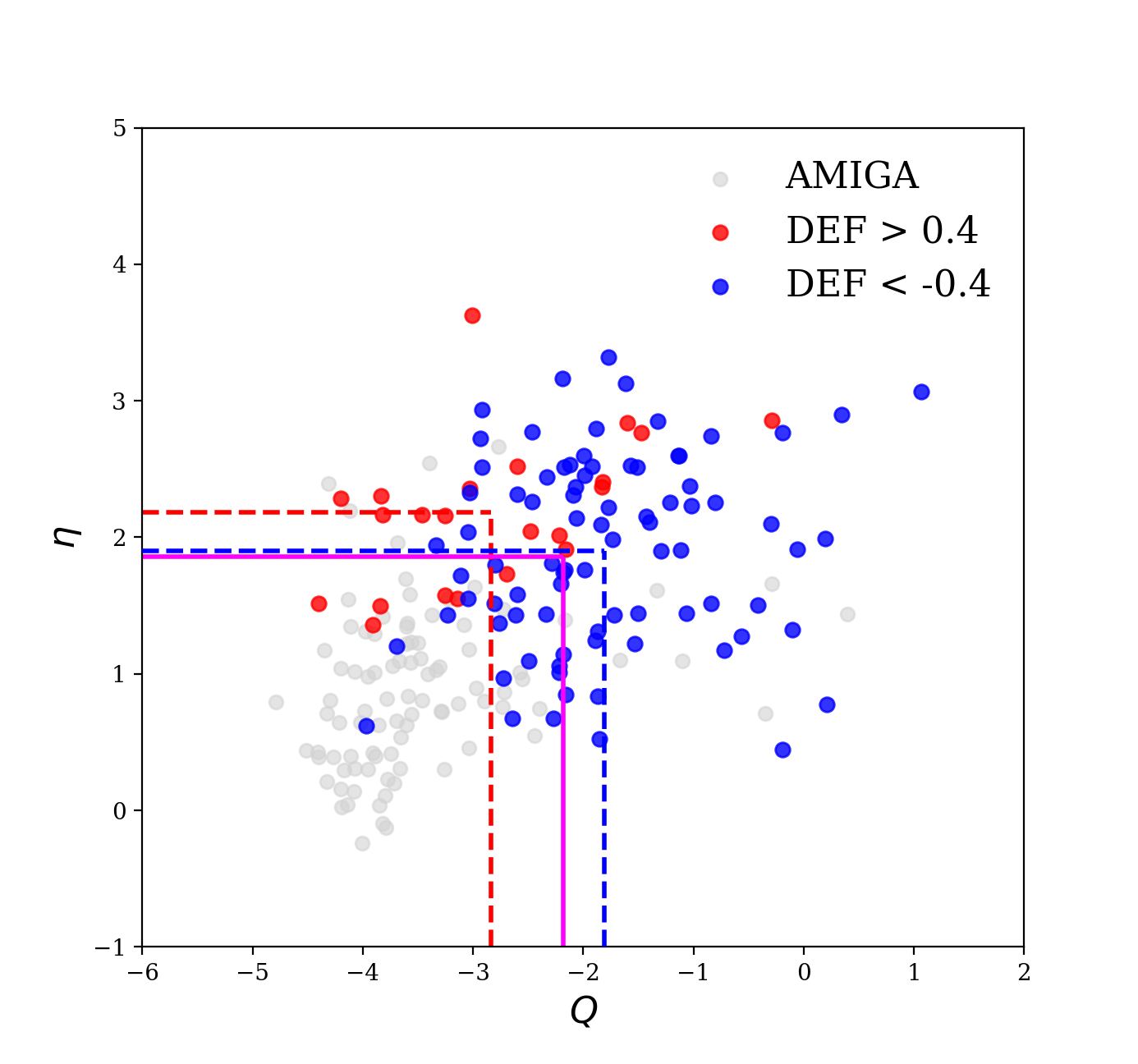}
    \caption{H{\textsc i~}deficiency on the $\eta-Q$ plane. Highly deficient (DEF>0.4) galaxies are shown in red, and galaxies with low deficiencies/excess H{\textsc i~}(DEF <-0.4) are shown in blue. AMIGA galaxies are shown in gray for reference. Red and blue dashed lines mark the mean $\eta$ and $Q$ values for the respective sub populations of high and low deficiency galaxies. Mean values for the full pair sample are marked by magenta lines.}
    \label{fig:highlowdef}
\end{figure}

\noindent Figure \ref{fig:highlowdef} highlights the location of both the most H{\textsc i~}deficient (here defined as DEF>0.4, i.e. 60$\%$ less H{\textsc i~}than expected given their stellar mass) galaxies in our pair sample (red points), as well as the most gas rich (DEF < -0.4, blue points), on the $\eta-Q$ plane. Visually we note a subtle separation between the two populations, with the most gas rich galaxies appearing to occupy a larger region of the $\eta-Q$ plane, spanning a broad range $\eta$ and $Q$ values, and in particular, extending out towards high $Q$ values. The most deficient galaxies, in contrast, rarely reach high $Q$ values. Quantitatively we note that, in general, larger tidal forces are associated with the gas-rich population compared to the gas-poor population, consistent with Figure \ref{fig:etaq2panels}.\\
 
\begin{figure*}
    \centering
    \includegraphics[scale = 0.65]{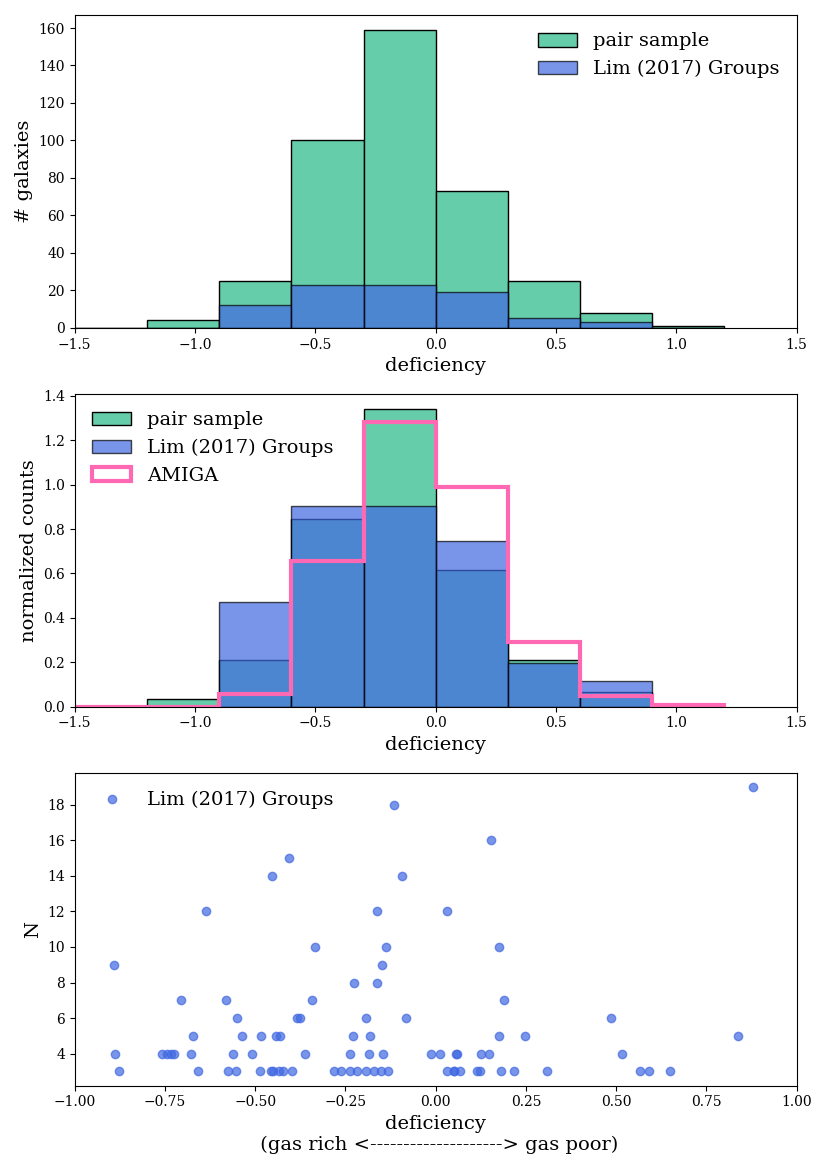}
    \caption{H{\textsc i~}deficiency in galaxy groups. Top panel: Distribution of H{\textsc i~}deficiency values for the pair sample (green) with the pairs contained within the \protect\cite{Lim2017} group catalogue shown in blue. Middle: Normalized H{\textsc i~}deficiency distributions for the pair, group, and isolated galaxy samples shown in green, blue, and pink respectively. Bottom panel: H{\textsc i~}deficiency of galaxies in groups as a function of the number group members, N. We see a broad range of deficiencies for groups of all sizes.}
    \label{fig:limvsdef}
\end{figure*}

\noindent In the top panel of Figure \ref{fig:limvsdef} we show separate deficiency distributions for the pair galaxies located in groups in blue and the remainder of the pair sample in green, and find no evidence to suggest that the group environment is specifically responsible for broadening the H{\textsc i~}deficiency distribution of the pair sample in either direction. In the middle panel of Figure \ref{fig:limvsdef} we compare the normalized H{\textsc i~}deficiency distributions of the pair (green), group (blue), and isolated (pink line) galaxy samples, and note that the group distribution appears shifted towards lower deficiency values (i.e. are more gas-rich) compared to both the pair and isolated samples. The bottom panel of Figure \ref{fig:limvsdef} displays the H{\textsc i~}deficiency for the group galaxies as a function of number of members in a group (N). Small groups (N<=4), which make up the majority of the group sample, exhibit a broad range in H{\textsc i~} deficiencies, and we see a tentative trend of larger groups (N>4) moving towards higher deficiency values as group membership increases, however, we note the limitations of small sample statistics in the large N regime.       

\subsection{Galaxy morphology and H{\textsc i~}deficiency}
\begin{figure}
    \centering
    \includegraphics[scale=0.5]{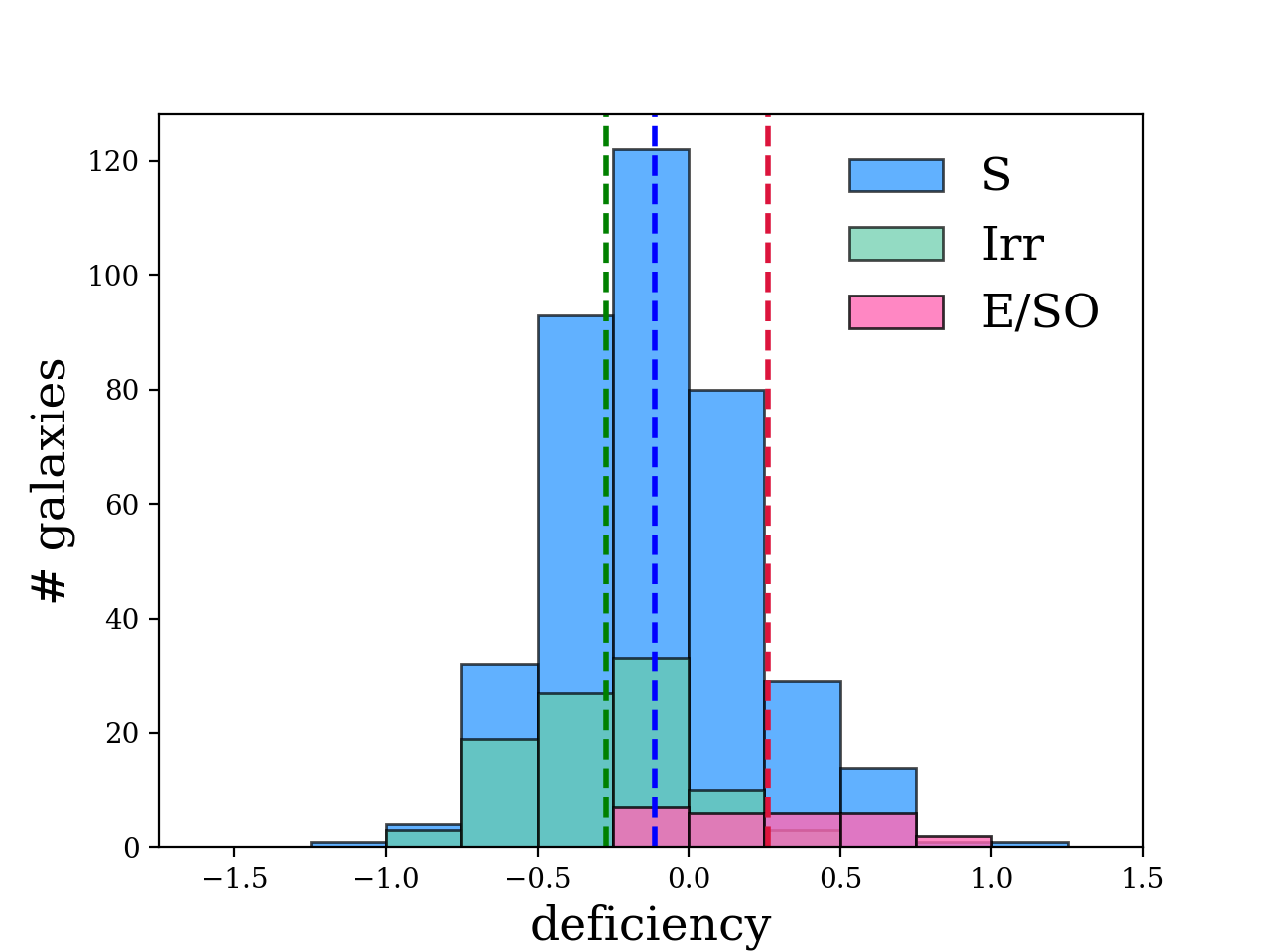}
    \caption{Deficiency distributions of the late-type (spirals (blue) and irregulars (green)), and early-type (elliptical/S0 (pink)) pair sub-samples. The corresponding mean deficiency values are shown as dashed vertical lines in blue, green, and pink respectively, highlighting the relative gas-richness of the spiral and irregular samples compared to the elliptical/S0 population.}
    \label{DEFmorph}
\end{figure}

\begin{figure*}
    \centering
    \includegraphics[scale=0.55]{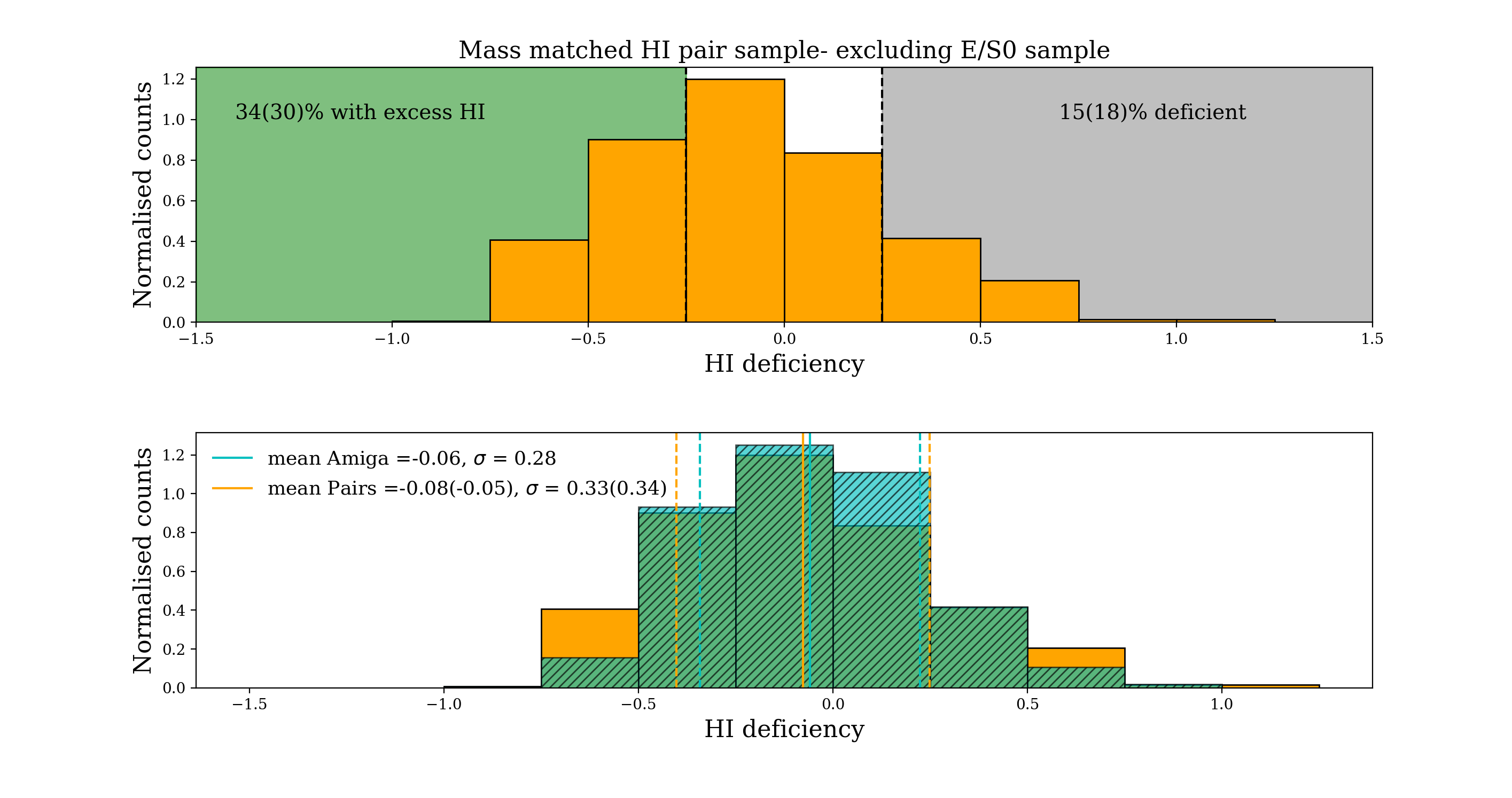}
    \caption{Top panel: Deficiency distribution of the pair sample excluding the E/S0 population, with the relative fractions of deficient galaxies and those with excess H{\textsc i~}shown as percentages on the plot. The full sample figures from \protect\cite{Bok2020} are indicated in brackets. Bottom panel: AMIGA-WISE H{\textsc i~}deficiency distribution (cyan) overlaid on the H{\textsc i~}pair deficiency distribution with the E/S0 subtracted. Cyan and orange vertical dashed lines mark the respective distribution widths, with the \protect\cite{Bok2020} values indicated in brackets.}
    \label{DEFmorphb}
\end{figure*}

In Figure \ref{DEFmorph} we show the pair member deficiency distributions of the spiral, irregular, and elliptical/S0 galaxy populations in blue, green, and pink respectively. The sample is spiral dominated as a consequence of our selection criteria, so the comparisons we make between the spiral and the elliptical/S0 galaxy sub-samples are not statistically significant, however we note that the elliptical/S0 deficiency distribution is shifted towards higher deficiencies, while the irregular galaxy sample galaxies are on average more gas-rich compared to both the spiral and elliptical/S0 populations. This result is not unexpected as later type galaxies are generally found to be more H{\textsc i-}rich compared to earlier types \citep{Solanes2001,Jones2018}. Since early-type galaxies are typically gas poor, it is unclear whether the high \hi deficiency values measured for these galaxies are driven at all by environment, or simply a consequence of morphology. Our sample of isolated galaxies in comparison is essentially devoid of elliptical/S0 galaxies (only $1\%$ are classified as E/S0), indicating that the broadening of the pair sample deficiency towards higher values is morphology-driven, at least in part. The strict AMIGA isolation criteria is strongly bias against early-type galaxies (which typically live in dense environments), although genuinely isolated early-types do exist \citep{Rampazzo2020A}. In Figure \ref{DEFmorphb} we test the impact of the early-type population on the pair deficiency distribution by excluding them from the pair sample, and note that their exclusion results in relatively fewer galaxies being classified as H{\textsc i~}deficient (15 $\%$ versus 18 $\%$) than was found previously for the full sample. The mean deficiency is additionally shifted to a slightly lower value of -0.09 when we exclude the E/S0 sample (compared to -0.05 for the full sample, i.e. less deficient on average), with a slightly smaller sigma value of $\sigma = 0.32$ (versus $\sigma = 0.34$ for the full sample). Controlling for morphology thus reduces the difference between the pair and AMIGA deficiency distributions, but does not eliminate it- the AMIGA sample still has a lower dispersion ($\sigma_{AMIGA}=0.28$.)

\subsection{Star formation efficiency}
\begin{figure*}
    \centering
    \includegraphics[scale = 0.55]{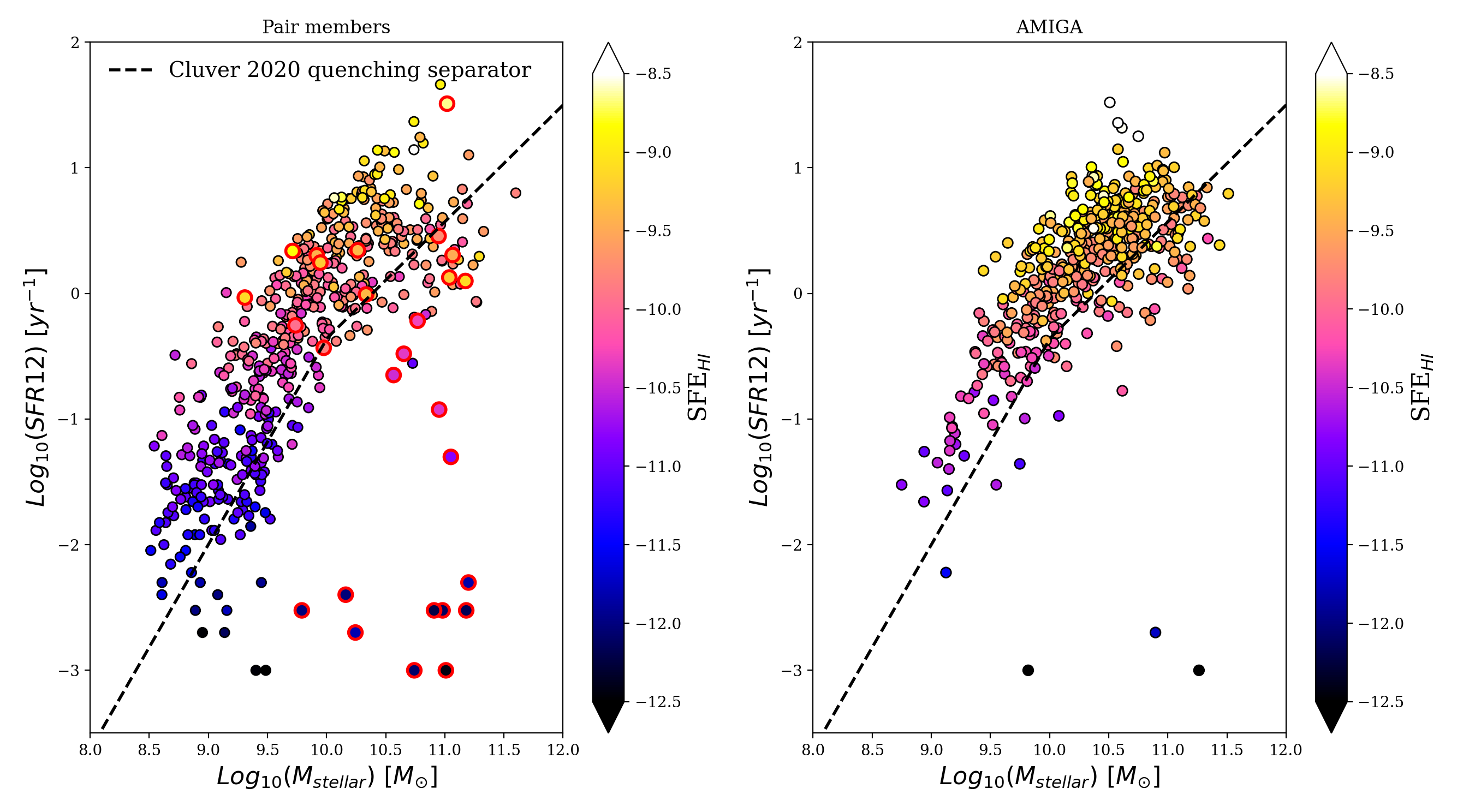}
    \caption{Pair (left) and isolated (right) SFMS colour coded by star formation efficiency (SFE(H{\textsc i})). early-type (E/S0) pair members are outlined in red. We see a general trend of SFE increasing along the SFMS. Galaxies with low SFRs appear to be similarly inefficient at forming stars regardless of their stellar mass.}
    \label{fig:MS_SFE}
\end{figure*}

\begin{figure*}
\includegraphics[scale =0.7]{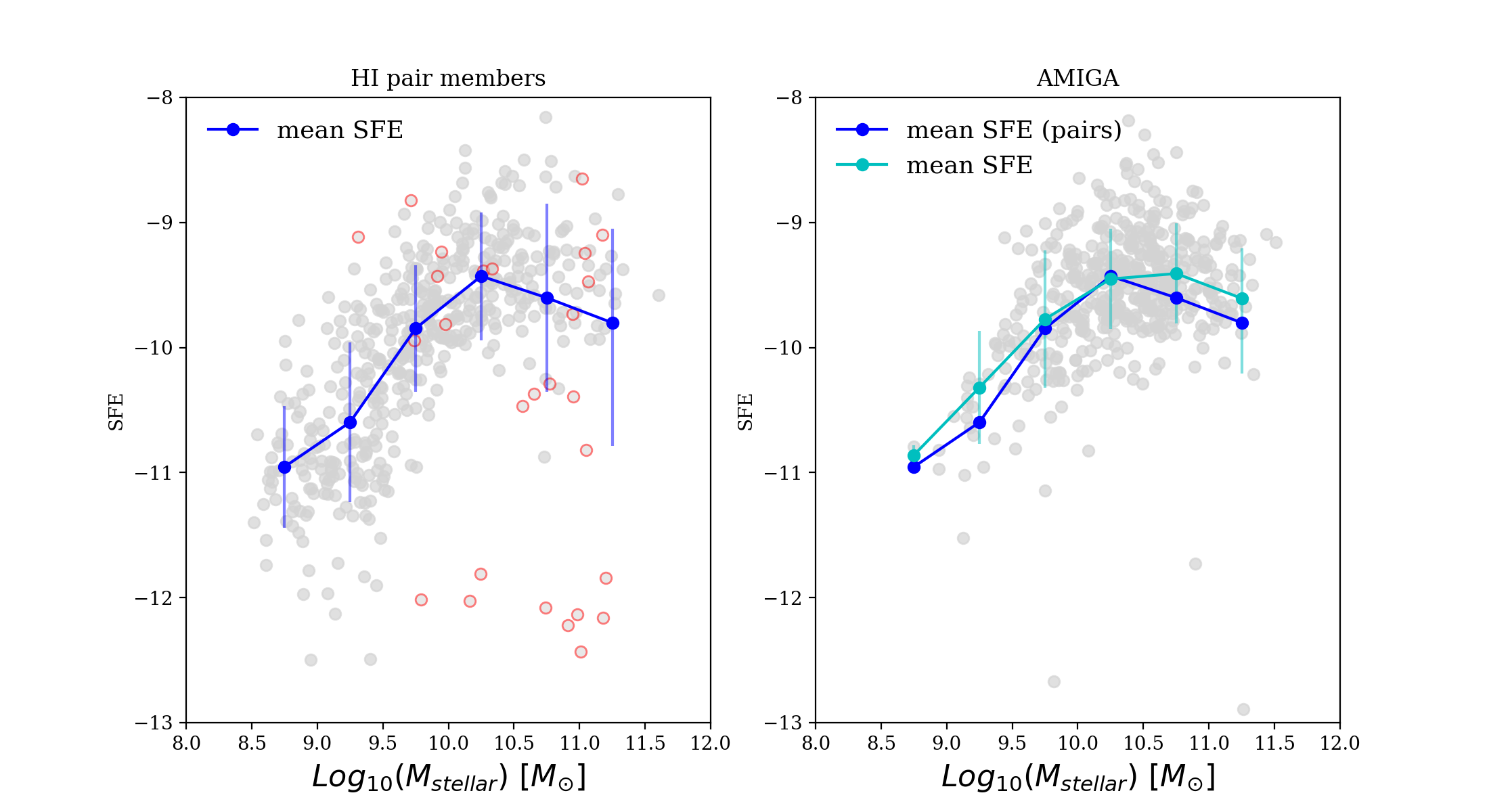}
\caption{Mean SFE(H{\textsc i}) per stellar mass bin (with standard errors depicted as vertical lines) for the pair (left panel) and isolated (right panel) galaxy samples illustrating a pronounced turnover at $\sim10^{10.25}$M$_{\odot}$ in both samples. We track the early-type pair members in red as they are expected to be passive, and indeed we do see a population of passive early-types occupying the lower right corner of the plot. We also note, however, the presence of early-type pair members that are actively and efficiently forming stars- potentially soon to join their passive kin.}
\label{meanSFE}
\end{figure*}

\begin{figure*}
    \centering
    \includegraphics[scale = 0.55]{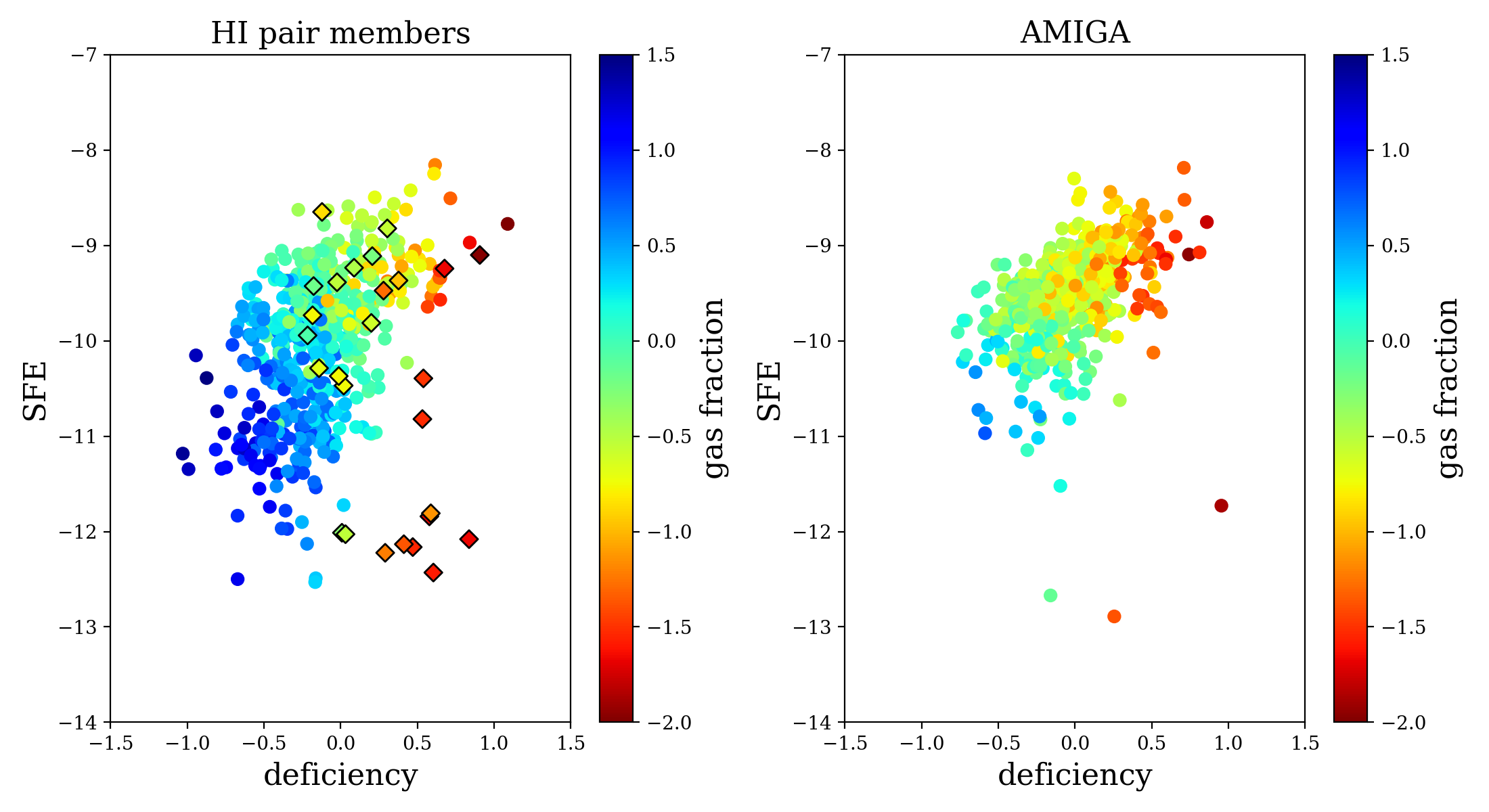}
    \caption{H{\textsc i~}deficiency as a function of SFE for the pair (left) and isolated (right) galaxy samples. Both plots are colour-coded by gas fraction. Efficient star formation appears to deplete galaxies of their gas reservoirs in both environments, with galaxies becoming increasingly deficient in H{\textsc i~}as SFE increases. In the pair sample we see SFE ultimately plummet for a small number of galaxies (bottom right corner of the plot), perhaps marking a later stage in the cycle, where previous high SF has depleted fuel supplies, and halted SF as a result. Indeed, if we look at morphology, these galaxies have been classified as early-type galaxies (shown here as diamonds), which are typically gas poor and no longer actively forming stars. We also point out the presence of a rather atypical population of early-type galaxies with relatively large gas fractions and high SFEs.}
    \label{fig:SFE_def}
\end{figure*}

While a supply of fuel is of course crucial for SF, how efficiently a galaxy is able to convert its gas into stars is equally important in driving galaxy location on the SFMS. Star formation efficiency (SFE) has been defined in the literature in many different ways, but in general refers to the star formation rate surface density per unit neutral or molecular gas surface density, i.e. SFE = $\Sigma$SFR/$\Sigma$H{\textsc i~}(/H$_2$) \citep{Leroy2008}. 
SFE is the inverse of gas depletion time (how long it would take for a galaxy to consume its gas reservoir given its current SF), and thus represents a combination of the real timescale for the conversion of gas into stars, and the fraction of gas that becomes stars \citep{Leroy2008}. Normalizing by $\Sigma$gas makes SFE particularly useful (more so than SFR) at identifying where conditions are conducive to SF \citep{Leroy2008}, and relates more directly to the mechanism of turning the interstellar gas into stars \citep{Inoue2000}. While SFE does not appear to correlate with galaxy morphology (for high mass spiral types; \cite{Devereux1991}), \cite{Young1999} found SFE(H$_2$) to decrease with increasing galaxy size for spirals across a broad range of environments (isolated, Virgo cluster spirals, spiral pairs, and field spirals). \cite{Young1999} suggests that the reduced SFE in large galaxies is a result of the greater shear associated with large disks, which is expected to increase the turbulent energy in molecular clouds (the location of SF), and thus disrupt SF . On the other hand, significantly enhanced SFEs are found in merging/interacting galaxies \citep{Young1986,Sanders1986,Solomon1988}. This result is invariant to the choice of SFR tracer (H$\alpha$ or FIR emission), and galaxy size \citep{Young1999}. One possible explanation for the enhanced SFEs observed in merging galaxies is the increased rate of molecular cloud collisions that occur during interactions \citep{Noguchi1986}, which would trigger enhanced SF in galaxies of all sizes \citep{Young1999}. \\\\
\cite{Schiminovich2010} calculated SFE(H{\textsc i}) for a representative sample of massive (>10$^{10}$M$_{\odot}$) galaxies from the \textit{GALEX} (\textit{Galaxy Evolution Explorer}) Arecibo SDSS (Sloan Digital Sky Survey) survey \citep{Catinella2010}, and found very little fluctuation in SFE across the sample. With no trends in stellar mass, stellar mass surface density, NUV-r colour and concentration observed for SFE, \cite{Schiminovich2010} propose that the regulation of SF in massive galaxies is likely driven primarily by processes that control the gas supply in galaxies (e.g. external processes or feedback mechanisms).\\\\
\noindent As we found in our previous paper \citep{Bok2020}, galaxies can have large gas reservoirs (high gas fractions) and very low SFRs. To investigate potential scenarios that could lead to these results we calculate the efficiency of SF for both our paired sample and our sample of isolated galaxies using the \cite{Schiminovich2010} definition of star formation efficiency: SFE = SFR/M$_{\textrm{H\textsc i}}$. In Figure \ref{fig:MS_SFE} we present the SFEs for both the paired (left) and isolated (right) galaxy samples as colour on the SFMS. We show the full galaxy samples here (not mass-matched), with the early-type pair members circled in red to acknowledge the role of morphology in driving trends. The relative fraction of paired and isolated galaxies in each stellar mass bin can be found in Table \ref{mstarbins}, which highlights the relative absence of low mass galaxies in the AMIGA sample. In both samples we observe a general trend of SFE increasing along the SFMS with increasing stellar mass and SFR. Migration off the SFMS and low SFEs appear in both samples, suggesting SFE is an important factor in quenching regardless of environment. This result is in agreement with the findings of \cite{Piotrowska2021}, who similarly observe reductions in both SFE and $f_{\rm{gas}}$ to be associated with quiescence, with SFE ultimately harnessing the greater predictive power of the two gas quantities in determining active/passive classifications. The increased frequency of galaxies migrating off the SFMS in the pair sample compared to the isolated sample can be largely attributed to early-type morphology (noted previously in Figure 2). The broad range in SFRs for the early-type galaxies at the turnover suggests that these galaxies are in the process of (rapid) transition. In Figure \ref{meanSFE} we plot the mean SFE per stellar mass bin for our pair (left) and isolated (right) samples, and observe a noticeable turnover at around SFMS $10^{10.25}$ \msol$~$ in both. This is consistent with the break in the stellar mass-halo mass relation \citep{Behroozi2013} corresponding to the mass scale where virial shock-heating of accreting gas begins to dominate. We compare the means directly in the right hand panel with the pair sample means shown in blue and the isolated sample in cyan, and find no significant difference between the pair and isolated trends. We tabulate the means and their errors per stellar mass bin in Table \ref{newtable}, which quantitatively demonstrates the statistical invariance of the pair SFE trend to the inclusion/exclusion of the early-type sample. \\\\ In Figure \ref{fig:SFE_def} we plot H{\textsc i~}deficiency as a function of SFE and colour code the data by gas fraction ($\rm{Log_{10}(M_{HI}/M_{star})}$. Here we see a subtle trend in both samples of increasing SFE with increasing deficiency, and decreasing gas fraction, suggesting \textit{efficient star formation depletes galaxies of their gas reservoirs.} This result is consistent with the work of \cite{Martinez-Badenes2012}, who found no clear trend with H{\textsc i~}or H$_{2}$ and deficiency for Hickson Compact Groups. A small population of low SFE, H{\textsc i~}deficient galaxies with depleted gas fractions perhaps marks a later stage in the cycle in which previous high SFE has depleted gas content, and in turn caused SFE to plummet. It is unsurprising that these galaxies have been classified as early-types by our visual classification scheme (shown as diamonds on the plot), however, it is notable that our gas-rich selection criterion does not preclude them from our study. The fact that there are few objects in the middle suggest this happens very rapidly. We also note that the AMIGA sample is missing high gas fraction galaxies relative to the pair sample. This is a selection effect arising from their strict isolation criteria, which is strongly biased against dwarfs (the original catalog used by \cite{Karachentseva1973} was not sensitive to faint objects, and as such was only able to detect nearby dwarfs. Assessing the isolation of nearby dwarfs, however, requires a prohibitively large search area for neighbours; \cite{Verley2007a}). Field dwarfs are naturally gas rich as they have only recently begun converting their gas to stars. Besides gas richness being associated with the early stages of galaxy formation/evolution, gas richness could also be the result of potential accretion, except for a galaxy to be accreting gas it is most likely not isolated, unless it's accreting directly from the cosmic web, not another galaxy. At the very least we do not expect to see the extreme case of accretion of gas from mergers in an isolated galaxy sample.

\begin{table*}
\centering
\caption{Fraction of paired and isolated galaxies per stellar mass bin.}
\begin{tabular}{@{}lllllll@{}}
\toprule
\textbf{Log(M$\star$) bins {[}M$_\odot${]}} & \textbf{8.5-9} & \textbf{9-9.5} & \textbf{9.5-10} & \textbf{10-10.5} & \textbf{10.5-11} & \textbf{\textgreater{}11} \\ \midrule
\textbf{Pair sample (\%)}          & 11.78          & 23.9           & 26.23           & 21.67            & 12.35            & 4.75                      \\
\textbf{Isolated sample (\%)}      & 0.62           & 5.61           & 19.95           & 35.13            & 30.35            & 8.31                      \\ \bottomrule
\end{tabular}
\label{mstarbins}
\end{table*}

\begin{table*}
\centering
\caption{Mean SFE per stellar mass bin for the pair sample (including and excluding the E/S0 sample), and the AMIGA sample.}
\begin{tabular}{@{}lllllll@{}}
\toprule
\textbf{Log(M$\star$) bins {[}M$_\odot${]}} & \textbf{8.5-9} & \textbf{9-9.5} & \textbf{9.5-10} & \textbf{10-10.5} & \textbf{10.5-11} & \textbf{\textgreater{}11} \\ \midrule
\textbf{Mean SFE pairs}          & -10.95 $\pm$ 0.49          & -10.60 $\pm$ 0.64          & -9.85 $\pm$ 0.51           & -9.43$\pm$ 0.51            & -9.60 $\pm$ 0.75           & -9.80 $\pm$ 0.98                      \\
\textbf{Mean SFE pairs (without early-types)}      & -10.95 $\pm$0.49          & -10.61$\pm$0.63           & -9.85 $\pm$0.47          & -9.38 $\pm$0.40           & -9.41 $\pm$0.47           & -9.47$\pm$0.32                     \\
\textbf{Mean SFE AMIGA}      & -10.86 $\pm$0.08          & -10.32$\pm$0.45           & -9.77 $\pm$0.55          & -9.45 $\pm$0.40           & -9.41 $\pm$0.40           & -9.61$\pm$0.60                     \\ \bottomrule
\end{tabular}
\label{newtable}
\end{table*}

\subsection{H{\textsc i~}profile asymmetry}
With the integrated H{\textsc i~}profiles readily available for our ALFALFA pair members, we maximise their utility by including a study of the profile asymmetries as a function of local environment, as well as H{\textsc i~}deficiency. \cite{Espada2011} found that the AMIGA sample of isolated galaxies present the smallest fraction of asymmetric H{\textsc i~}profiles compared with any other yet studied, and proposed that it can serve as a baseline for studying asymmetry rates in other environments. Indeed, in a previous paper \citep{Bok2019} we explored the possibility of tracing merger activity via asymmetries measured in the H{\textsc i~}profiles of galaxies in close pairs using a simple flux ratio to quantify asymmetry. Our results demonstrated that the close pair environment does indeed enhance profile asymmetry, and we proposed that asymmetries measured in the high asymmetry regime ($A_{c}$>1.26) could, in turn, be used to infer potential merger activity in close pairs. In our 2019 paper we focused exclusively on ALFALFA galaxies with a single close optical (not H{\textsc i}) companion (i.e. isolated H{\textsc i-}optical pairs); here in this updated work, we use the pair sample of \cite{Bok2020} in which a single close H{\textsc i~}companion is a minimum requirement. Systems with additional companions (H{\textsc i~}and otherwise) are not excluded. Consequently, the pair sample of \cite{Bok2020} spans a broad range of local environments, allowing us to study H{\textsc i~}profile asymmetry in this more inclusive local environment bracket, and explore the plausible proposition that density of local environment and the degree of measured H{\textsc i~}profile asymmetry positively correlate. We calculate asymmetry following the method described in Section 4 of \cite{Bok2019}, selecting only good quality profiles (flagged as such by the ALFALFA team), with SNR >10. We show in \cite{Bok2019} that for SNR >= 10, the maximum uncertainty in $A_{c}$ is at 5\%, which is calculated using Monte-Carlo methods (described in section 4.1), and therefore put forward that SNR is unlikely impacting our asymmetry results in any significant way. \\

\begin{figure*}
    \centering
    \includegraphics[scale=0.55]{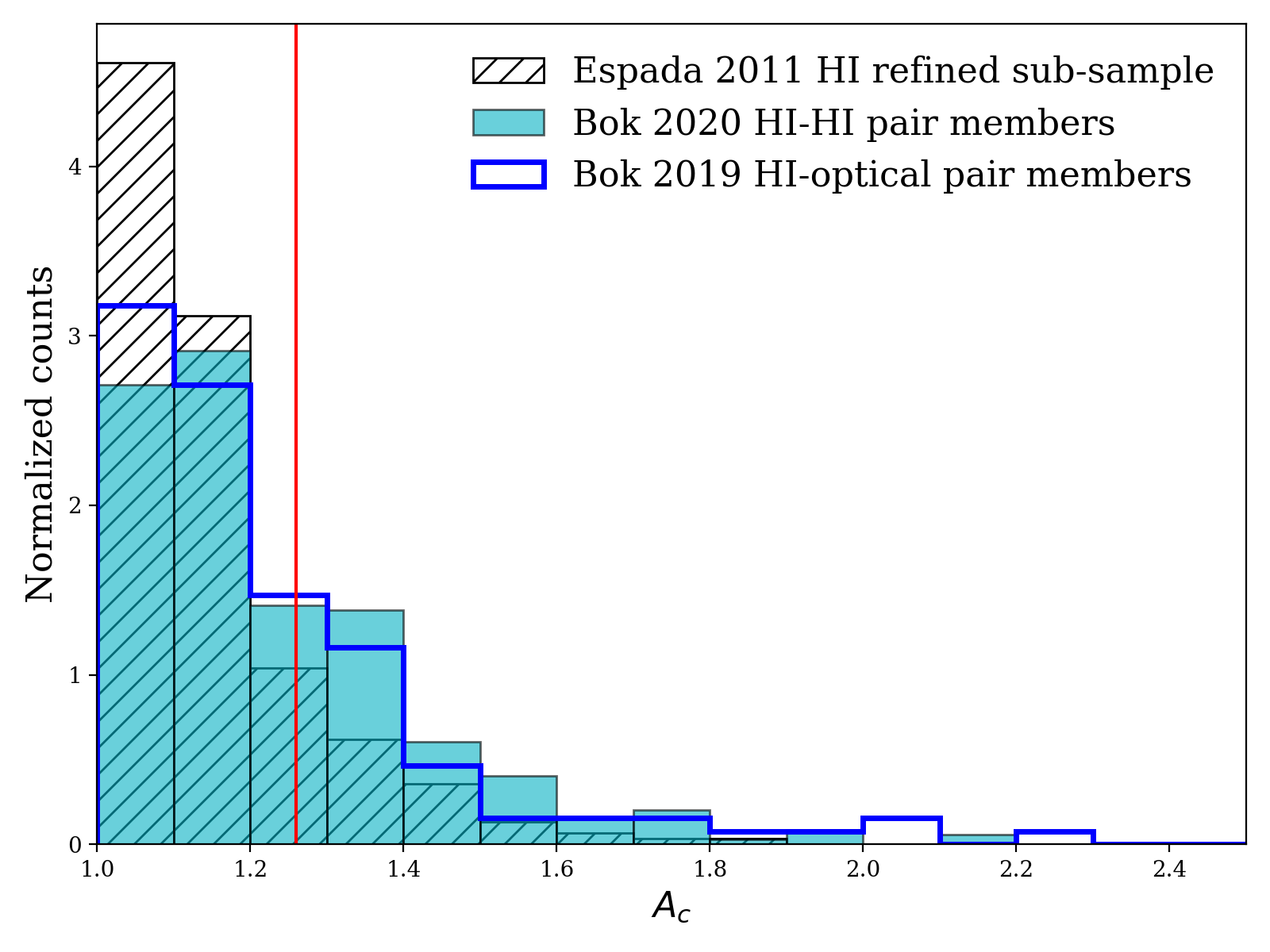}
    \caption{Distribution of the Bok 2020 H{\textsc i~}pair sample asymmetry measurements (cyan), the Bok 2019 H{\textsc i}-optical pair sample asymmetry measurements (blue line), and the Espada 2011 asymmetry measurements for the H{\textsc i~}refined sub-sample of AMIGA galaxies (hatched histogram). The red line marks the threshold for profile symmetry at $A_{c}$ = 1.26, above which profiles are considered asymmetric.}
    \label{fig:asymdistributions}
\end{figure*}

\begin{table}
\centering
\caption{H{\textsc i~}profile asymmetry measurements from the literature}
\begin{tabular}{llll}
\hline
Galaxy sample &size & $A_{c}$\textgreater 1.26 & error \\ \hline
H{\textsc i~}refined subsample (Espada et al., 2011) & 166         & 9$\%$                & 2.2$\%$        \\
Haynes et al., 1988                        & 104         & 9$\%$                & 2.8$\%$        \\
Matthews et al., 1998                      & 30          & 17$\%$               & 6.8$\%$        \\
H{\textsc i~}isolated sample (Bok et al. ,2019)      & 304         & 18$\%$               & 2.2$\%$        \\
H{\textsc i~}optical pair sample (Bok et al., 2019)  & 304         & 27$\%$               & 2.6$\%$        \\
\textbf{H{\textsc i~}pair members (Bok et al., 2020)} & \textbf{253} & \textbf{32$\%$} & \textbf{2.6$\%$} \\ \hline
\end{tabular}
\label{asymtable}
\end{table}

\noindent In Figure \ref{fig:asymdistributions} we contrast the pair sample asymmetry distributions of \cite{Bok2020} (cyan) and \cite{Bok2019} (blue line) with asymmetry measurements from \cite{Espada2011} for the H{\textsc i~}refined sub-sample of AMIGA galaxies (hatched histogram). \cite{Espada2011} described the asymmetry distribution of their isolated galaxy sample as a half Gaussian centred on 1 (representing perfect symmetry) with a width $\sigma = 0.13$. In recent studies by \cite{Scott2018} and \cite{Bok2019}, galaxies with $A_{c}$>  1.26 (corresponding to the 2$\sigma$ width of the isolated distribution) are considered significantly asymmetric. Qualitatively we note that the \cite{Bok2019,Bok2020} samples follow similar distributions, and are similarly different from the \cite{Espada2011} sample of isolated galaxy asymmetries in that both exhibit extended tails towards high asymmetries. In Table \ref{asymtable} we compare directly with the literature, and note a significantly larger fraction of the \cite{Bok2020} pair sample lie in the $A_{c} > 1.26$ bins ($32\%$) compared to all previous samples in less densely populated environments: $9\%$ for the isolated galaxies (\cite{Espada2011,Haynes1988}), $17\%$ for field galaxies (\cite{Matthews1998}), and $27\%$ for isolated pairs (\cite{Bok2019}). \\

\begin{figure}
    \centering
    \includegraphics[scale = 0.57]{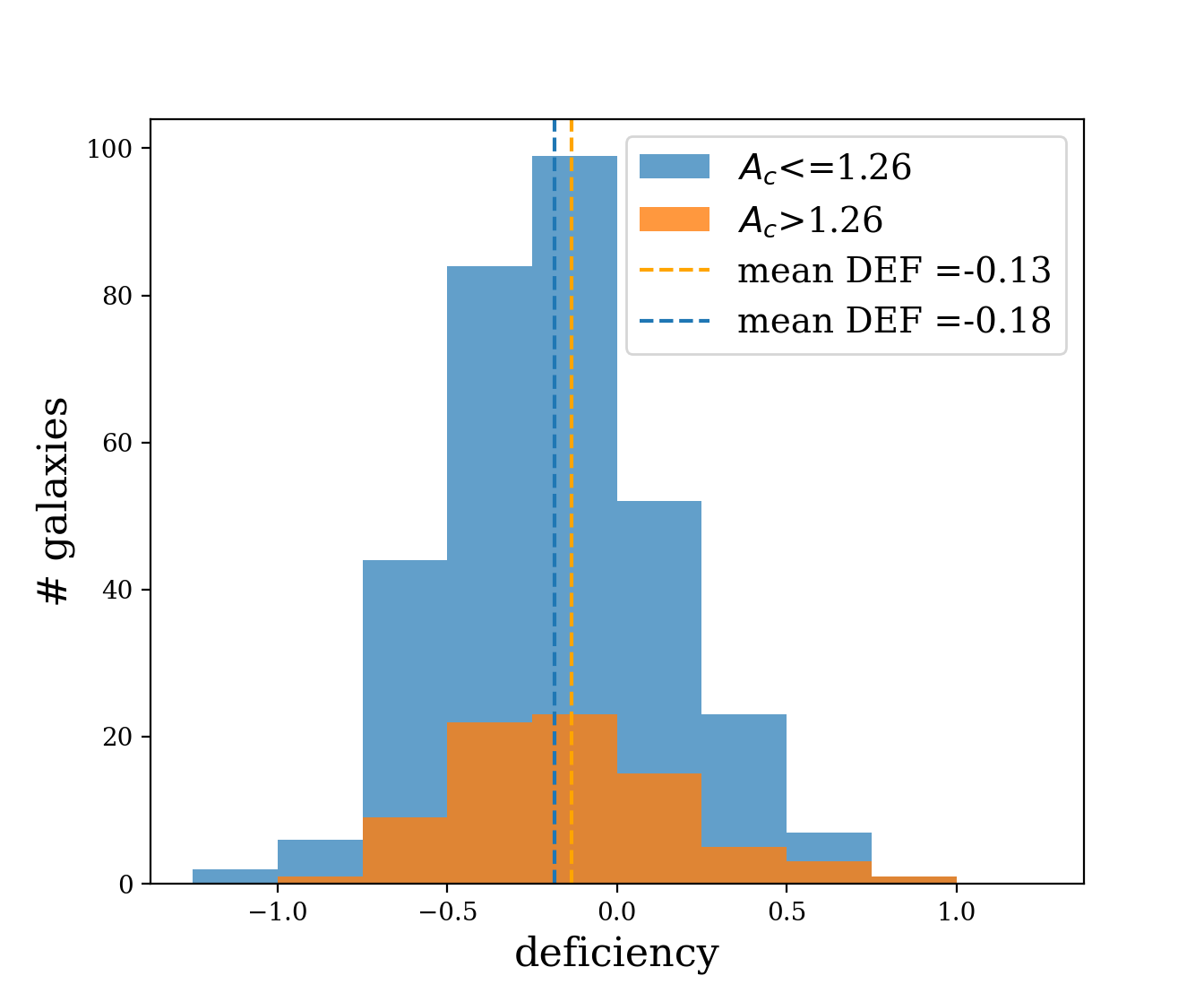}
    \caption{Here we contrast the asymmetric ($A_{c}$>1.26) deficiency distribution (orange) with the symmetric ($A_{c}$<=1.26) deficiency distribution (blue).}
    \label{acdef}
\end{figure}

\noindent Figure \ref{acdef} displays the H{\textsc i~} deficiency distributions of the asymmetric ($A_{c}$>1.26) and symmetric ($A_{c}$<=1.26) populations separately as orange and blue histograms respectively. This plot highlights the relatively small size of the asymmetric population compared to the symmetric population (32$\%$ asymmetric versus 68$\%$ symmetric), as well as how similarly shaped the asymmetric and symmetric deficiency distributions are. Quantitatively we note that the mean deficiency of the asymmetric population (orange dashed line) is similar to the symmetric deficiency mean (blue dashed line), and both means (-0.13 and -0.18 for the respective asymmetric and symmetric samples) lie in the `normal' gas content range (-0.25<DEF<0.25). A statistical significance test for the equality of the two distributions is provided by the two-sample Kolmogorov-Smirnov test, which confirms that the asymmetric and symmetric deficiency distributions are not significantly different (D = 0.09, p-value = 0.72).\\

\begin{figure*}
    \centering
    \includegraphics[scale = 0.6]{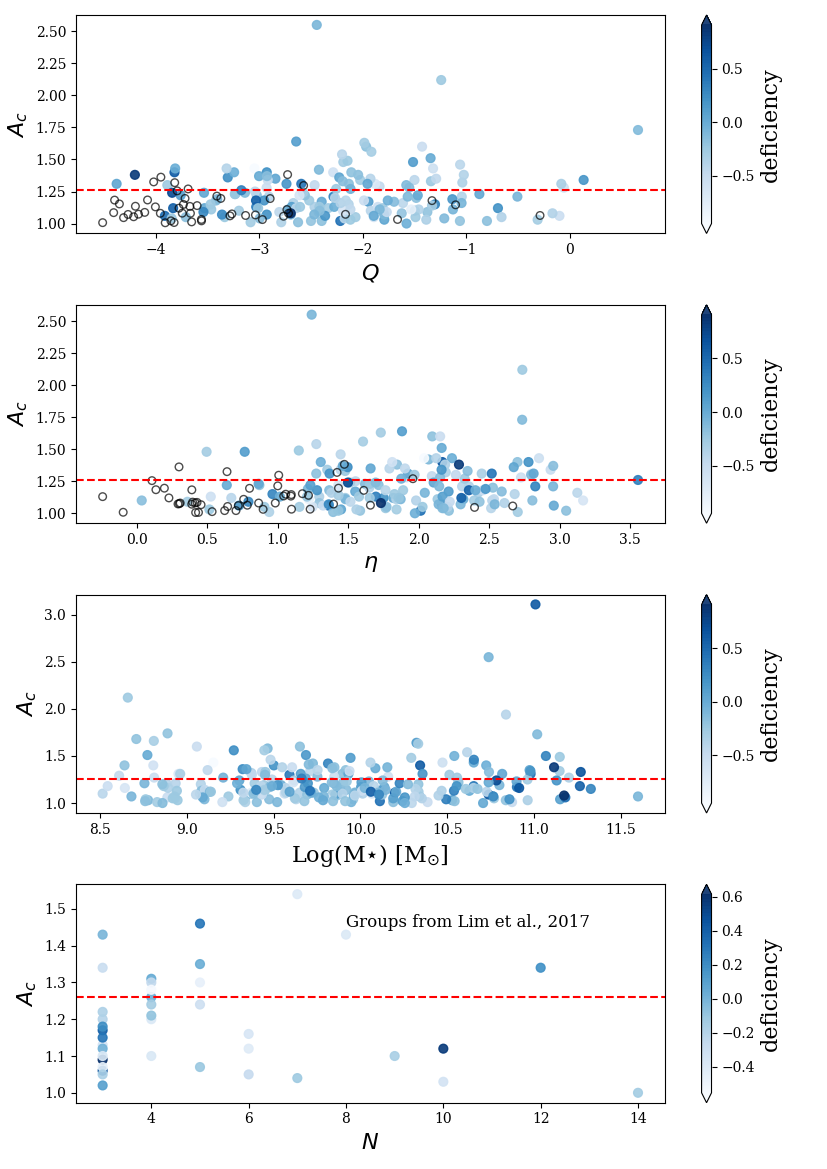}
    \caption{From top to bottom: Asymmetry ($A_{c}$) as a function of tidal strength (Q), local number density ($\eta$), stellar mass, and number of group members (N) for the \protect\cite{Lim2017} groups in the pair sample, colour-coded by H{\textsc i~}deficiency. The red horizontal dashed line in each plot corresponds to $A_{c}$=  1.26, marking the baseline for asymmetry (profiles with $A_{c}$> 1.26 are considered significantly asymmetric). Empty circles (outlined in black) show the location of the AMIGA galaxies for reference in panels 1 and 2. While there are no apparent trends with asymmetry and Q, $\eta$, or stellar mass, we note that the largest asymmetries (those with $A_{c}$> 1.75) have Q values strictly greater than -2.5. The majority of the group galaxies (shown in the forth panel) do not measure significant profile asymmetries, and of those that do, the degree of asymmetry does not appear to depend on N.}
    \label{ac_ALL}
\end{figure*}

\begin{table*}
\centering
\caption{Fraction of asymmetric galaxies in bins of Q, $\eta$, H{\textsc i~} deficiency, Log(M$\star$)}
\begin{tabular}{|c|c|c|c|c|c|c|}
\hline
\textbf{Q bins}                                                                            & \textbf{{[}-5,-4)}                                   & \textbf{{[}-4,-3)}                                   & \textbf{{[}-3,-2)}                                   & \textbf{{[}-2,-1)}                                   & \textbf{{[}-1,0)}                                    & \textbf{\textgreater{}0}                             \\ \hline
\textbf{\begin{tabular}[c]{@{}c@{}}$A_{c}$\textgreater{}1.26 (\%)\\ (\# galaxies)\end{tabular}} & \begin{tabular}[c]{@{}c@{}}2.53\\ (2)\end{tabular}  & \begin{tabular}[c]{@{}c@{}}10.13\\ (8)\end{tabular} & \begin{tabular}[c]{@{}c@{}}29.11\\ (23)\end{tabular} & \begin{tabular}[c]{@{}c@{}}25.31\\ (20)\end{tabular} & \begin{tabular}[c]{@{}c@{}}2.53\\ (2)\end{tabular}  & \begin{tabular}[c]{@{}c@{}}30.34\\ (24)\end{tabular}    \\ \hline
\textbf{$\eta$ bins}                                                                       & \textbf{{[}0,0.5)}                                   & \textbf{{[}0.5,1)}                                   & \textbf{{[}1,1.5)}                                   & \textbf{{[}1.5,2)}                                   & \textbf{{[}2,2.5)}                                   & \textbf{\textgreater{}2.5}                           \\ \hline
\textbf{\begin{tabular}[c]{@{}c@{}}$A_{c}$\textgreater{}1.26 (\%)\\ (\# galaxies)\end{tabular}} & \begin{tabular}[c]{@{}c@{}}29.11\\ (23)\end{tabular} & \begin{tabular}[c]{@{}c@{}}1.27\\ (1)\end{tabular}   & \begin{tabular}[c]{@{}c@{}}15.18\\ (12)\end{tabular} & \begin{tabular}[c]{@{}c@{}}15.18\\ (12)\end{tabular} & \begin{tabular}[c]{@{}c@{}}22.78\\ (18)\end{tabular} & \begin{tabular}[c]{@{}c@{}}16.46\\ (13)\end{tabular} \\ \hline
\textbf{Deficiency bins}                                                                   &\textbf{\textless{}-0.5}                             & \textbf{{[}-0.5,-0.25)}                              & \textbf{{[}-0.25,0)}                                 & \textbf{{[}0,0.25)}                                  & \textbf{{[}0.25,0.5)}                                & \textbf{\textgreater{}0.5}                           \\ \hline
\textbf{\begin{tabular}[c]{@{}c@{}}$A_c$\textgreater{}1.26 (\%)\\ (\# galaxies)\end{tabular}} & \begin{tabular}[c]{@{}c@{}}12.66\\ (10)\end{tabular} & \begin{tabular}[c]{@{}c@{}}27.85\\ (22)\end{tabular} & \begin{tabular}[c]{@{}c@{}}29.11\\ (23)\end{tabular} & \begin{tabular}[c]{@{}c@{}}18.99\\ (15)\end{tabular} & \begin{tabular}[c]{@{}c@{}}6.33\\ (5)\end{tabular}  & \begin{tabular}[c]{@{}c@{}}5.06\\ (4)\end{tabular}  \\ \hline

\textbf{Log(M$\star$) bins {[}M$_\odot${]}}                                                                   &\textbf{{[}8.5,9)}                             & \textbf{{[}9,9.5)}                              & \textbf{{[}9.5,10)}                                 & \textbf{{[}10,10.5)}                                  & \textbf{{[}10.5,11)}                                & \textbf{\textgreater{}11}                           \\ \hline
\textbf{\begin{tabular}[c]{@{}c@{}}$A_c$\textgreater{}1.26 (\%)\\ (\# galaxies)\end{tabular}} & \begin{tabular}[c]{@{}c@{}}13.92\\ (11)\end{tabular} & \begin{tabular}[c]{@{}c@{}}18.99\\ (15)\end{tabular} & \begin{tabular}[c]{@{}c@{}}26.58\\ (21)\end{tabular} & \begin{tabular}[c]{@{}c@{}}12.66\\ (10)\end{tabular} & \begin{tabular}[c]{@{}c@{}}17.72\\ (14)\end{tabular}  & \begin{tabular}[c]{@{}c@{}}10.13\\ (8)\end{tabular}  \\ \hline

\end{tabular}
\label{binsofQ}
\end{table*}

\noindent In the top two panels of Figure \ref{ac_ALL} we look at Q and $\eta$ individually as a function of asymmetry, with H{\textsc i~}deficiency additionally encoded as colour. The dashed red horizontal line in each panel corresponds to $A_{c}$=  1.26, signifying the baseline value for asymmetry. We find no clear trends between asymmetry and either environment parameter ($\eta$ or Q). In Table \ref{binsofQ} we show the frequency of asymmetry ($\%$ of galaxies with $A_{c}$> 1.26) in bins of both Q, $\eta$, and H{\textsc i~}deficiency, and again find no indication of a correlation between $A_{c}$ and the AMIGA environment parameters, or H{\textsc i~}content. If we look exclusively at the highest asymmetries in the pair sample (those with $A_{c}$> 1.75), we note that their Q values are strictly larger than -2.5, and that they correspond in general to galaxies that are either gas-rich (DEF < -0.25) or have normal H{\textsc i~} content (-0.25<DEF<0.25) according to their H{\textsc i~}deficiency values, with only 2/7 measuring a positive deficiency value.\\\\ In the fourth panel of Figure \ref{ac_ALL} we show profile asymmetries measured for the \cite{Lim2017} group galaxies as a function of their group size (N), as well as H{\textsc i~}deficiency. We note first that the majority of H{\textsc i~}gas (in pairs) lives in small (low N) groups, and that only a relatively small fraction of these group galaxies measure significant H{\textsc i~}profile asymmetries (13$\%$ with $A_{c}$> 1.26). Of the galaxies that are asymmetric in their profiles, group size (N) does not appear to influence the degree of asymmetry, although we acknowledge small numbers in the high N regime. For small groups in particular (N<=4), we observe a broad range of asymmetries, suggestive of active transformation. We additionally look at $A_{c}$ as a function of stellar mass in panel 4, which demonstrates these quantities to be uncorrelated. We quote the fraction of asymmetric galaxies per stellar mass bin in the last row of Table \ref{binsofQ}, and confirm that stellar mass cannot be used to estimate the frequency of asymmetry in our pair sample.\\\\
\noindent Table \ref{highAC} lists the most asymmetric ($A_{c}$>1.75) galaxies in our pair sample, together with their Q values, deficiency values, and their local environments as per visual inspection of their optical images. This table demonstrates how similarly high profile asymmetries can arise from a range of different configurations. For example, the gas-poor merger, AGC6965, and the gas rich merger, AGC201011, have equally asymmetric profiles (both with $A_{c}$ = 1.94). Similar can be said for both pairs and groups. 

\begin{table}
\centering
\caption{Q and DEF values for the high asymmetry ($A_c$ \textgreater 1.75) population, with notes on the local environment as per visual inspection of the optical SDSS cutouts  indicated in the last column. Red indicates a galaxy is gas-poor (i.e. DEF\textgreater{}0.25), and blue indicates gas-richness (DEF\textless{}-0.25).}
\label{highAC}
\begin{tabular}{@{}ccccl@{}}
\toprule
\textbf{AGC \#} & \textbf{$A_c$} & \textbf{Q} & \textbf{DEF}                 & \textbf{Environment note} \\ \midrule
6965            & 1.94        & -0.29      & {\color[HTML]{FE0000} 0.53}  & merger                    \\
201011          & 1.94        & --         & {\color[HTML]{3531FF} -0.40} & merger                    \\
215143          & 2.12        & -1.24      & {\color[HTML]{3531FF} -0.25} & pair                      \\
230069          & 1.91        & -2.51      & 0.13                         & pair                      \\
240367          & 2.55        & -2.44      & -0.08                        & potential group           \\
1286            & 3.11        & --         & {\color[HTML]{FE0000} 0.6}   & potential group           \\
330779          & 1.77        & --         & 0.2                          & potential group           \\ \bottomrule
\end{tabular}
\end{table}


\section{Discussion}
By invoking the use of the local number density ($\eta$) and tidal influence (Q) parameters, galaxy morphologies, and H{\textsc i~}profile asymmetries to further characterize our pair sample galaxies, we start to differentiate between the high and low H{\textsc i~}deficiency tails of the pair sample deficiency distribution in \cite{Bok2020} as arising from different sub-populations in the sample, and associated with different dynamics. In Figures \ref{fig:etaq2panels} and \ref{fig:highlowdef} we see that the low deficiency (gas-rich) tail of the distribution corresponds to higher tidal influence (Q) values and lower number densities ($\eta$) in general compared to the H{\textsc i~}deficient population, however note that these differences are not statistically significant when we compare their means (Q$_{\rm{mean}}$ = -1.9 and $\eta_{\rm{mean}}$ = 1.9  for the gas rich sample (DEF<-0.25), versus Q$_{\rm{mean}}$ = -2.8 and $\eta_{\rm{mean}}$ = 2.1 for the gas poor sample (DEF>0.25)). This suggests a local environment in which the target galaxies have a few (low $\eta$) relatively large (high Q) neighbours. \\\\Simulations of merging galaxies show that the merger process is capable of elevating the cold-dense gas content in both primaries and secondaries \citep{Moreno2020} by way of gravitational torques, and observations of merging/recently merged galaxies show that these galaxies are indeed gas-rich compared to control samples \citep{Ellison2018}. Our results indicate the possible presence of a sub-sample of currently merging/recently merged galaxies in our pair sample, who are gas rich (i.e. with low H{\textsc i~}deficiencies) as a result of their merger activity, which we infer from their large Q values (Q$_{\rm{mean}}$ = -1.9). We note that the deficiency distribution of the irregular galaxy population is also shifted to lower deficiencies compared to both the dominant spiral and small elliptical/S0 galaxy populations (see Figure \ref{DEFmorph}), strengthening the argument for a merging sub-sample if we attribute the disturbed, irregular morphologies to merger activity, which is plausible given the coincidence of high Q values. The link between merger activity and disturbed optical morphologies is indeed well documented in the literature, e.g. \cite{Patton2005, DePropis2007,Ellison2010}. The largest H{\textsc i~}profile asymmetries ($A_c$>1.75) in our pair sample are also associated with Q values strictly larger than -2.5 (see top panel of Figure \ref{ac_ALL}), and we put forward that the link is likely causal. \cite{Sulentic2006} found a similar link between high Q values and optically distorted galaxy morphologies. \cite{Janowiecki2017} found that central galaxies in small groups, with $M_{\star}<10^{10.2}$ \msol, have larger gas fractions on average compared to isolated galaxies of the same stellar mass, as well as enhanced SF activity- this signal disappears at larger stellar masses. \cite{Janowiecki2017} attribute the relative gas-richness of these low mass centrals to their location at slightly richer parts of the cosmic web, where the availability of H{\textsc i~}is elevated compared to more isolated locations, or H{\textsc i~}rich minor mergers. A large fraction of our pair sample falls in this low mass category, and it is therefore plausible that a richer cosmic web location is contributing to their elevated H{\textsc i~}content. \\\\

\noindent On the other hand, the high deficiency tail of the pair sample distribution appears to arise from a different set of circumstances. As we see in Figures \ref{fig:etaq2panels} and \ref{fig:highlowdef}, the most deficient galaxies in the sample live on average in more densely populated environments relative to the low deficiency (gas-rich) sample ($\eta_{\rm{mean}}$=2.1 versus $\eta_{\rm{mean}}$ = 1.9, respectively). They are also subject to smaller tidal forces on average (Q$_{\rm{mean}}$ = -2.8 versus -1.9). This suggests a local environment of relatively many, but smaller companions, compared to the gas-rich sample $\eta$-Q environment, more akin to a group environment than a merger-pair environment. If these galaxies are indeed living in gravitationally-bound groups, their deficiency in H{\textsc i~}can be understood as a consequence of the group environment, which \cite{HessWilcots2013} demonstrate will produce increasingly H{\textsc i~}deficient central cores as optical membership increases. Interestingly, the confirmed groups in our sample (those that are part of the \cite{Lim2017} galaxy group catalogue), do not appear to be particularly deficient in H{\textsc i~}compared to the rest of the sample, nor does their group membership (N) seem to trend with deficiency. Looking to morphology again (Figure \ref{DEFmorph}), the small population of early-type galaxies in the pair sample live almost exclusively in the positive deficiency wing of the deficiency distribution, with a mean deficiency significantly higher than both the spiral and irregular galaxy populations (DEF$_{\rm{mean}}$ = 0.26, -0.11, and -0.27 for the E/S0, spiral, and irregular galaxy populations respectively). \cite{Solanes2001} also found early-type galaxies to be H{\textsc i~}deficient relative to later-types in their study. The presence of early-type galaxies in our pair sample (and relative absence of early-type galaxies in our isolated sample) suggests that morphology is likely contributing to the broadening of the pair sample deficiency distribution towards high H{\textsc i~}deficiencies relative to the isolated sample deficiency distribution. \\\\
\noindent In addition to a broadened deficiency distribution, our pair sample also has a larger low SF/quiescent population relative to the isolated galaxy sample (see Figure 3.6 in \cite{Bok2020}). Upon examination of the H{\textsc i~}content of these galaxies, two sub-populations emerged: a low stellar mass/gas-rich population, and a high mass/gas-poor population. The distinctly different gas and mass properties of these two populations suggest different pathways to their quenching. In Figure \ref{fig:MS_SFE} we colour code the SFMS by star formation efficiency (SFE) and find that the low SFR galaxies in both the pair and isolated samples all have similarly low SFEs, implicating SFE as an important pathway to quenching regardless of environment, stellar mass, and gas content. \\\\
\begin{figure}
    \centering
    \includegraphics[scale = 0.56]{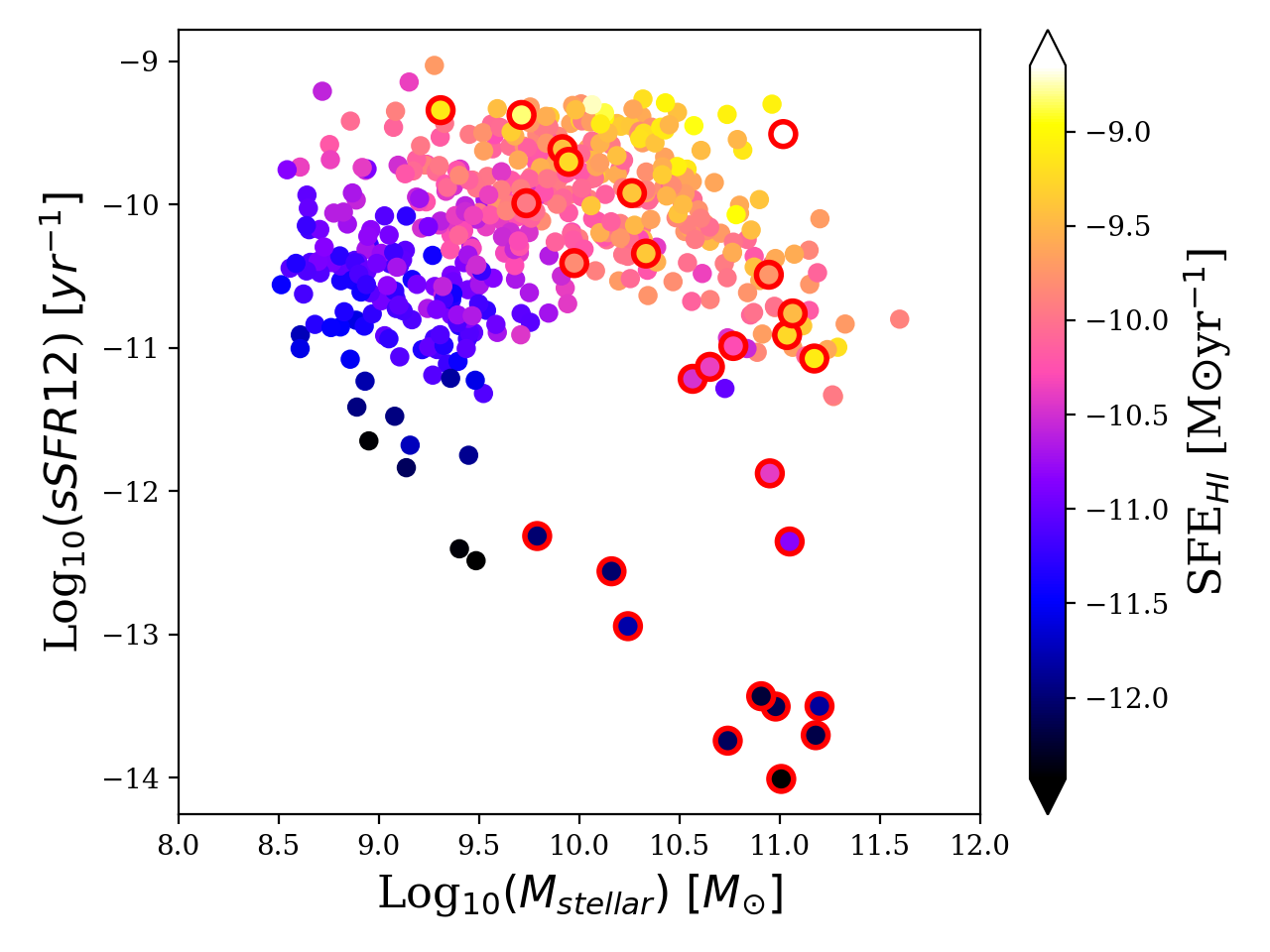}
    \caption{Specific star formation (log10(SFR/M$\star$)) versus stellar mass, colour coded by SFE (SFR/M$_{{\rm{\hi}}}$), with early-types circled in red. Here we note that our low SFR/SFE populations, previously identified as either low mass and gas rich or high mass and gas poor, can also be separated in sSFR. Both populations have low sSFRs with respect to the main SF sample, however the low mass population is relatively more active than the high mass population.}
    \label{fig:specificSFR}
\end{figure}
\noindent In Figure \ref{fig:specificSFR} we plot the specific SFRs (sSFR = SFR/M$_{\star}$) for our pair galaxies as a function of their stellar mass, and indicate their SFE as colour. Here we note that the low SFR, and low SFE populations previously categorized as either low mass and gas rich, or high mass and gas poor, are additionally separate in their specific star formation rates. While both populations have low sSFRs relative to the main SF population, the low mass galaxies have higher sSFRs (on average) compared to the high mass group (i.e. both populations are under-performing for their weight in terms of SF, and this is more pronounced in the high mass group. Low SFRs and sSFRs in the high mass/gas-poor population can plausibly be explained as driven by the consumption fuel (hence gas-poor) into stars (high stellar mass). With gas reserves depleted, continued SF is unsustainable, and these galaxies can be described as (fast approaching) fully quenched. The low mass population, on the other hand, with their extensive gas reservoirs, may continue to build up their stellar mass via slow and inefficient SF for many years to come, and might then build enough mass to become more efficient star formers. Low SFEs in this population are potentially the result of merger activity, which \cite{Moreno2020} showed could undermine SF in galaxies with cold-dense gas reservoirs. The ultimate fate of these galaxies may simply be slow death by fuel consumption, or perhaps external processes may intervene and alter their SF journeys, just as external processes may have already undermined their SF by impeding SFE. The fact that we see an increased frequency of low SFE galaxies in the pair sample relative to our isolated sample is potentially the result of merger induced turbulence in the pair sample, which \cite{Ellison2018} propose can reduce SFE, and thereby hinder effective SF. Angular momentum, which is large in large gas disks (high gas fraction galaxies) \citep{Obreschkow2016}, may also be contributing to the low SF in the gas rich population by providing resistance to collapse along the axis of rotation. \\\\
Our H{\textsc i~} profile analysis revealed our pair sample to have a significantly higher fraction of asymmetric profiles (32$\%$ with $A_{c}$> 1.26) compared to previous samples in the literature, where a similar flux ratio was used to quantify profile asymmetry in less densely populated (isolated and field) environments (i.e. 9$\%$,9$\%$,17$\%$,18$\%$ with $A_{c}$> 1.26 in the \cite{Espada2011,Haynes1988,Matthews1998} and \cite{Bok2019} samples respectively). We attribute this strong signal in asymmetry in our pair sample to the close pair environment, in which tidal interactions are likely at play. Our pair sample also has a higher fraction of asymmetric profiles compared to the pair sample used in \cite{Bok2019}, in which 27$\%$ (versus 32$\%$) of the H{\textsc i~}profiles were classified as asymmetric. The key differences between the two pair samples are the gas-rich requirements of the pair members used in this study (M$_{\rm{HI}}$>10$^9$ M$_{\odot}$), as well as the allowance of additional companions (more densely populated environments) in the 2020 sample. Examining our profile asymmetries as function of the AMIGA environment parameters reveals no clear trend of increasing asymmetry with density of environment ($\eta$), or tidal strength (Q). We also find no evidence to suggest that group size (N) influences the degree of asymmetry when we specifically look at the asymmetry measurements for the \cite{Lim2017} groups in our pair sample as a function of N. The majority of the group galaxies, in fact, do not measure a significant profile asymmetry (only 13$\%$ with $A_{c}$> 1.26). Our results show no signal of asymmetry being enhanced in groups relative to close pairs, nor between more densely populated environments (as per the AMIGA environment parameters) and close pairs. Similar asymmetry values are found for a broad range of environments, and conversely, particular environments (e.g. small groups) produce a variety of $A_{c}$ values, ranging from symmetric to highly symmetric. While we certainly see an enhanced frequency of asymmetry in close pair galaxies relative to isolated galaxies, from which we may infer potential merger activity between close companions, the asymmetry value alone cannot be used to distinguish the close pair environment from more densely populated environments.  \\

\noindent With regards to gas-content, the deficiency distributions of the asymmetric and symmetric populations are qualitatively similar in shape (see Figure \ref{acdef}). Using the two-sample Kolmogorov-Smirnov test to statistically compare the asymmetric and symmetric deficiency distributions, we find that the samples are indeed not significantly different (D = 0.013, p-value = 0.091). \cite{Watts2021} present similar findings in their recent study of the relationship between H{\textsc i~} profile asymmetry and gas content. Using the same flux ratio to quantify H{\textsc i~} profile asymmetry as the one used in this paper, they find no systematic differences in the H{\textsc i~} content of asymmetric and symmetric galaxies. \textit{Our findings here, and those of \cite{Watts2021}, indicate the limitations of H{\textsc i~} profile asymmetries in communicating their origins.} Such an endeavour may be better suited to  cosmological-hydrodynamical galaxy evolution simulations, which \cite{Watts2020} have already shown capable of providing valuable insight into the potential drivers of asymmetry in H{\textsc i~} profiles, specifically how a particular asymmetry can arise from multiple physical processes.



\section{Summary}
\begin{itemize}
    \item We compute the AMIGA environment parameters $\eta$ and $Q$ for our pair member galaxies and find that our pair members occupy a broad region on the $\eta-Q$ plane, and distinctly separate from the isolated galaxy population (AMIGA galaxies). We find a tentative trend of decreasing Q with increasing deficiency, and no trend between $\eta$ and deficiency. 
    \item We visually classify our pair member morphologies and find the sample to consist largely of late-type galaxies (75$\%$ spirals and 19$\%$ irregulars), with early-types accounting for $5\%$ of the sample. Our visual classification scheme is in good agreement with the MIR morphology diagnostics presented for the pair members in \cite{Bok2020} ({\it{WISE}} colour-colour diagram and B/T radial profile decomposition measurements).
    \item We examine morphology on the SFMS and show that the low SF galaxies in the pair sample (below the SFMS) are dominated by early-type morphologies, and that these galaxies are shifted towards higher H{\textsc i~}deficiencies. These results, together with the absence of early-types in the AMIGA sample, suggest morphology is responsible, at least in part, for the broadening of the pair sample deficiency distribution towards large \hi deficiencies relative to the AMIGA sample.
    \item{We present the star-formation efficiencies for both samples on the SFMS and find that regardless of environment, stellar mass, and H{\textsc i~}content, galaxies that have migrated off the SFMS are extremely inefficient at forming stars, with low specific star formation rates. In general SFE increases along the SFMS in both samples, and when we look at the mean SFEs per stellar mass bin we find a turnover in SFE at $\sim10^{10.25}$ \msol (just before, and perhaps driving the SFMS turnover at $\sim10^{10.5}$ \msol).}
    \item{We compute the H{\textsc i~}profile asymmetries of the pair sample using a standard flux ratio, and find a significantly higher rate of high profile asymmetries (with $A_{c}$>  1.26) compared to samples in less dense environments in the literature. However, we find no evidence of the degree of profile asymmetry increasing with density of environment within our pair sample when we examine asymmetry as a function of $\eta$, Q, and group size (N) for the galaxy groups in our sample. We also find no trends between profile asymmetry and H{\textsc i~}deficiency.}
\end{itemize}

\noindent In this paper we invoke various galaxy properties derived from optical, radio, and MIR data to investigate potential pathways for the observed trends we find for H{\textsc i~}deficiency and galaxy location on the SFMS as a function of environment. While we provide plausible scenarios with the data in hand, inferring potential mechanisms for accretion, gas removal, and quenching, as well as merger activity, from quantities such as H{\textsc i~}deficiency, H{\textsc i~}profile asymmetry, galaxy morphology (optical and MIR), and the AMIGA isolation parameters ($\eta$ and Q), the next generation of radio surveys with SKA1-MID are expected to provide direct evidence of accretion, as well as disentangle various gas removal mechanisms, by probing the intergalactic medium (IGM)/intragroup medium in unprecedented detail. State of the art simulations that incorporate more accurate astrophysical processes into their routines, i.e. \cite{Moreno2020}'s FIRE2 simulations, will provide important context for these next-generation observations. 

\section*{Acknowledgements}

This work is based on the research supported in part by the National Research Foundation of South Africa (Grant Numbers UID: 101099 and 111745). JB additionally acknowledges
support from the DST-NRF Professional Development Programme (PDP), and the University of Cape Town. MC is a recipient of an Australian Research Council Future Fellowship (project number FT170100273) funded by the Australian Government. THJ acknowledges funding from the National Research Foundation under the Research Career Advancement and South African
Research Chair Initiative programs, respectively.
We acknowledge the work of the entire ALFALFA team for observing, flagging and performing signal extraction. MGJ is supported by a Juan de la Cierva formaci\'{o}n fellowship (FJCI-2016-29685) from the Spanish Ministerio de Ciencia, Innovacion y Universidades (MCIU). MGJ also acknowledges support from the grants AYA2015-65973-C3-1- R (MINECO/FEDER, UE) and RTI2018-096228-B-C31 (MCIU).  This work has been supported by the State Agency for Research of the Spanish MCIU through the ‘Centro de Excelencia Severo Ochoa’ award to the Instituto de Astrof\'{i}sica de Andaluc\'{ı}a (SEV-2017-0709).
This publication makes use of data products from the Wide-field
Infrared Survey Explorer, which is a joint project of the University of California, Los Angeles, and the Jet Propulsion
Laboratory/California Institute of Technology, funded by the National Aeronautics and Space Administration.
This work also utilizes data from Arecibo Legacy Fast ALFA (ALFALFA) survey data set obtained with the Arecibo L-band Feed Array (ALFA) on the Arecibo 305m telescope. Arecibo Observatory is part of the National Astronomy and Ionosphere Center, which is operated by Cornell University under Cooperative Agreement with the U.S. National Science Foundation. Funding for the SDSS and SDSS-II has been provided by the Alfred P. Sloan Foundation, the Participating Institutions, the National Science Foundation, the U.S. Department of Energy, the National Aeronautics and Space Administration, the Japanese Monbukagakusho, the Max Planck Society, and the Higher Education Funding Council for England. The SDSS Web Site is http://www.sdss.org/. In addition, we make use of data from the Sloan Digital Sky Survey (SDSS DR7). The SDSS is managed by the Astrophysical Research Consortium for the Participating Institutions. The Participating Institutions are the American Museum of Natural History, Astrophysical Institute Potsdam, University of Basel, University of Cambridge, Case Western Reserve University, University of Chicago, Drexel University, Fermilab, the Institute for Advanced Study, the Japan Participation Group, Johns Hopkins University, the Joint Institute for Nuclear Astrophysics, the Kavli Institute for Particle Astrophysics and Cosmology, the Korean Scientist Group, the Chinese Academy of Sciences (LAMOST), Los Alamos National Laboratory, the Max-Planck-Institute for Astronomy (MPIA), the Max-Planck-Institute for Astrophysics (MPA), New Mexico State University, Ohio State University, University of Pittsburgh, University of Portsmouth, Princeton University, the United States Naval Observatory, and the University of Washington.

\section*{Data Availability}

The data underlying in this paper will be made available by request from J. Bok (email: jamie@saao.ac.za).



\bibliographystyle{mnras}
\bibliography{ref.bib} 




\section*{Appendix: Early-type sample}
\begin{figure*}

     \begin{subfigure}[b]{0.24\textwidth}
         \centering
         \includegraphics[width=\textwidth]{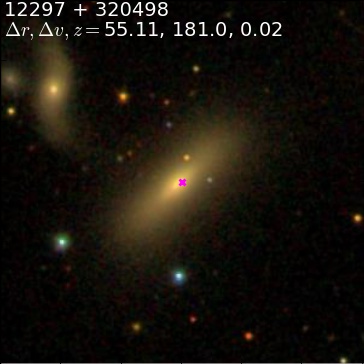}
     \end{subfigure}
     \hfill
     \begin{subfigure}[b]{0.24\textwidth}
         \centering
         \includegraphics[width=\textwidth]{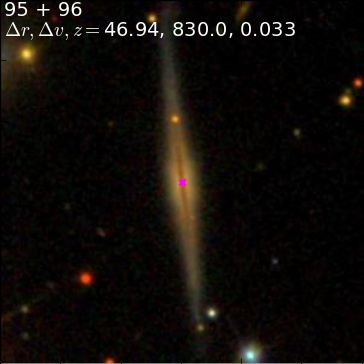}
     \end{subfigure}
     \hfill
     \begin{subfigure}[b]{0.24\textwidth}
         \centering
         \includegraphics[width=\textwidth]{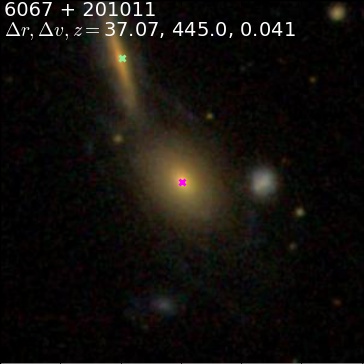}
     \end{subfigure}
     \hfill
     \begin{subfigure}[b]{0.24\textwidth}
         \centering
         \includegraphics[width=\textwidth]{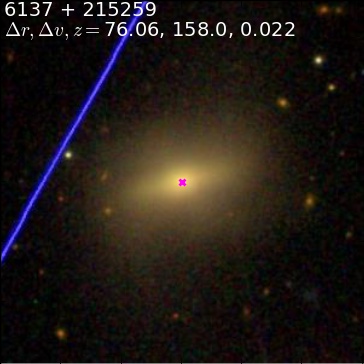}
     \end{subfigure}
     \hfill
     \begin{subfigure}[b]{0.24\textwidth}
         \centering
         \includegraphics[width=\textwidth]{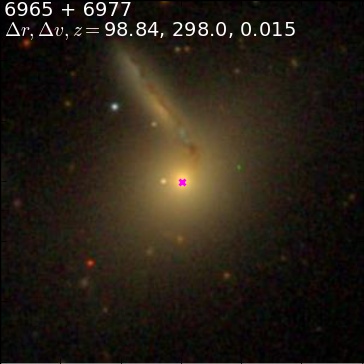}
     \end{subfigure}
     \hfill
     \begin{subfigure}[b]{0.24\textwidth}
         \centering
         \includegraphics[width=\textwidth]{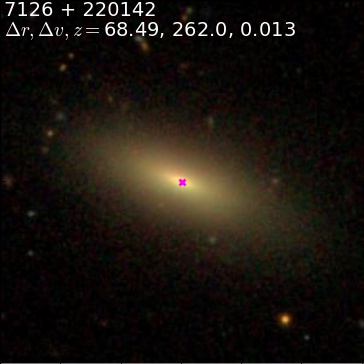}
     \end{subfigure}
     \hfill
     \begin{subfigure}[b]{0.24\textwidth}
         \centering
         \includegraphics[width=\textwidth]{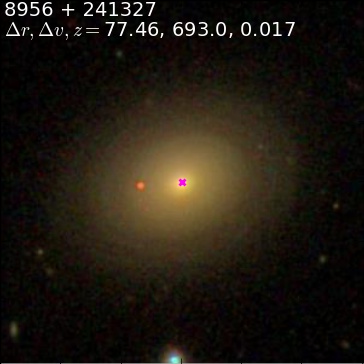}
     \end{subfigure}
     \hfill
     \begin{subfigure}[b]{0.24\textwidth}
         \centering
         \includegraphics[width=\textwidth]{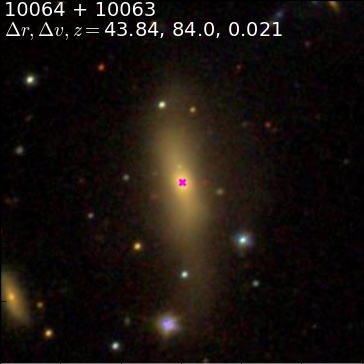}
     \end{subfigure}
     \hfill
          \centering
     \begin{subfigure}[b]{0.24\textwidth}
         \centering
         \includegraphics[width=\textwidth]{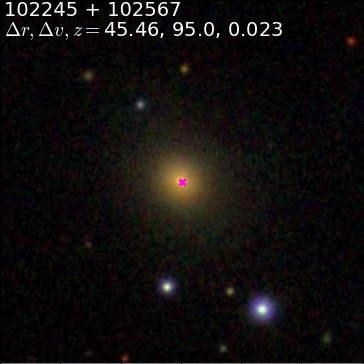}
     \end{subfigure}
     \hfill
     \begin{subfigure}[b]{0.24\textwidth}
         \centering
         \includegraphics[width=\textwidth]{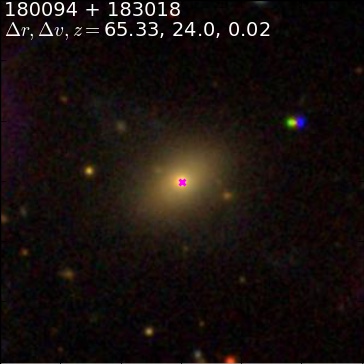}
     \end{subfigure}
     \hfill
     \begin{subfigure}[b]{0.24\textwidth}
         \centering
         \includegraphics[width=\textwidth]{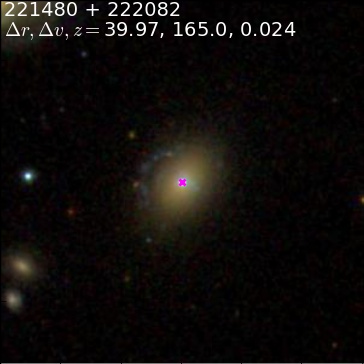}
     \end{subfigure}
     \hfill
     \begin{subfigure}[b]{0.24\textwidth}
         \centering
         \includegraphics[width=\textwidth]{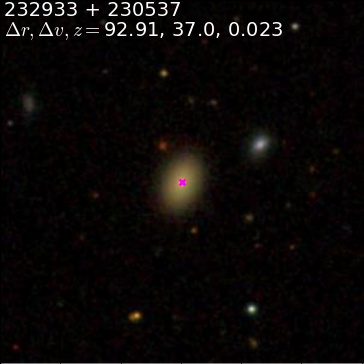}
     \end{subfigure}
     \hfill

     \begin{subfigure}[b]{0.24\textwidth}
         \centering
         \includegraphics[width=\textwidth]{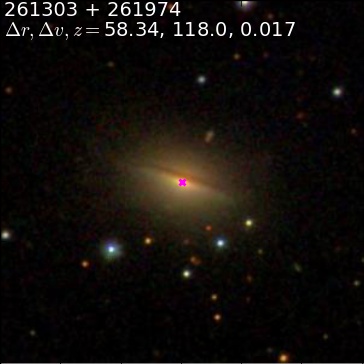}
     \end{subfigure}
     \hfill
     \begin{subfigure}[b]{0.24\textwidth}
         \centering
         \includegraphics[width=\textwidth]{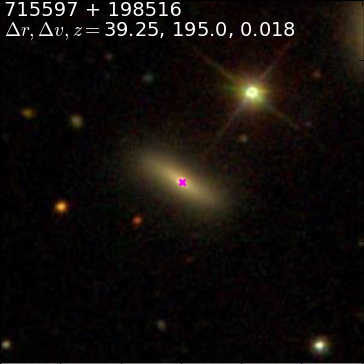}
     \end{subfigure}
     \hfill
     \begin{subfigure}[b]{0.24\textwidth}
         \centering
         \includegraphics[width=\textwidth]{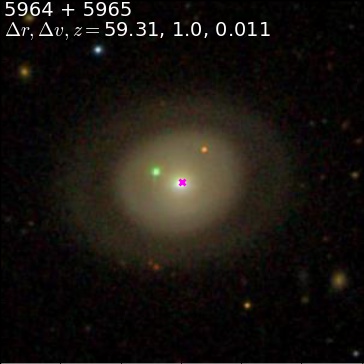}
     \end{subfigure}
     \hfill
     \begin{subfigure}[b]{0.24\textwidth}
         \centering
         \includegraphics[width=\textwidth]{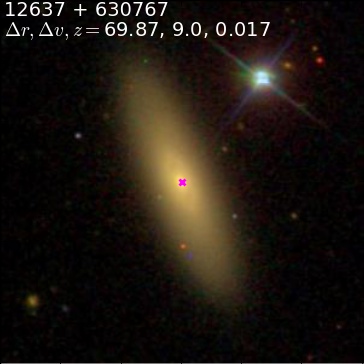}
     \end{subfigure}
     \hfill
     \begin{subfigure}[b]{0.24\textwidth}
         \centering
         \includegraphics[width=\textwidth]{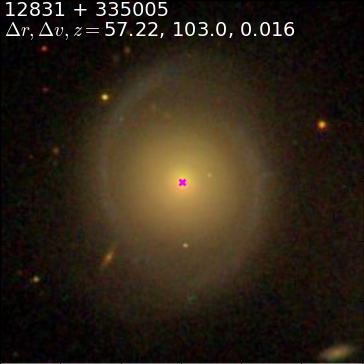}
     \end{subfigure}
     \hfill
     \begin{subfigure}[b]{0.24\textwidth}
         \centering
         \includegraphics[width=\textwidth]{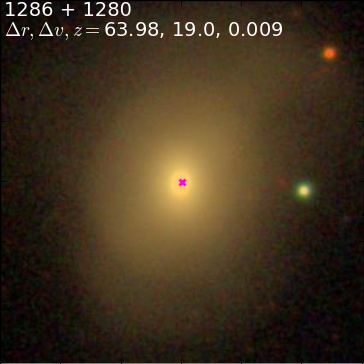}
     \end{subfigure}
     \hfill
     \begin{subfigure}[b]{0.24\textwidth}
         \centering
         \includegraphics[width=\textwidth]{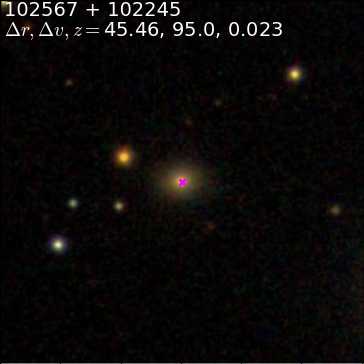}
     \end{subfigure}
     \hfill
     \begin{subfigure}[b]{0.24\textwidth}
         \centering
         \includegraphics[width=\textwidth]{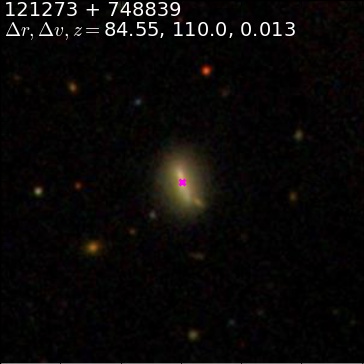}
     \end{subfigure}
     \hfill 
\end{figure*}
\begin{figure*}
\ContinuedFloat
     \begin{subfigure}[b]{0.24\textwidth}

         \includegraphics[width=\textwidth]{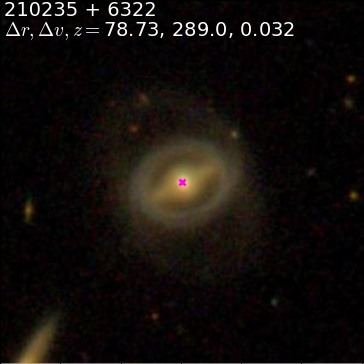}
     \end{subfigure}
     \hfill
     \begin{subfigure}[b]{0.24\textwidth}
         \centering
         \includegraphics[width=\textwidth]{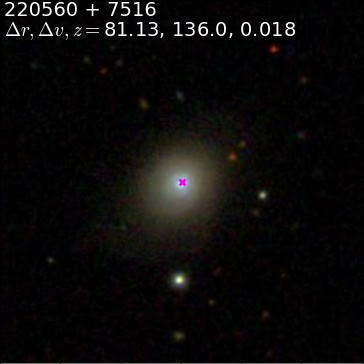}
     \end{subfigure}
     \hfill
     \begin{subfigure}[b]{0.24\textwidth}
         \centering
         \includegraphics[width=\textwidth]{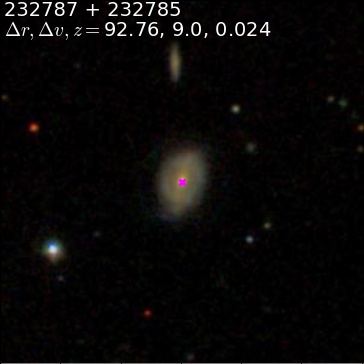}
     \end{subfigure}
     \hfill
     \begin{subfigure}[b]{0.24\textwidth}
         \centering
         \includegraphics[width=\textwidth]{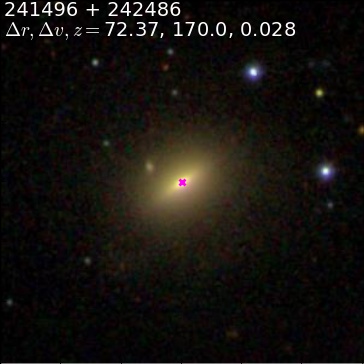}
     \end{subfigure}
     \hfill
     \begin{subfigure}[b]{0.24\textwidth}

         \includegraphics[width=\textwidth]{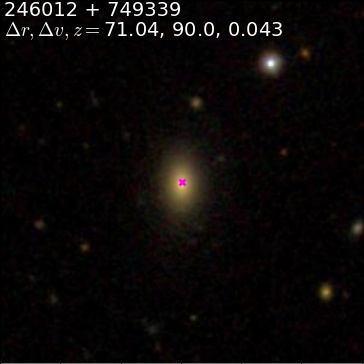}
     \end{subfigure}
     \hfill
     \begin{subfigure}[b]{0.24\textwidth}

         \includegraphics[width=\textwidth]{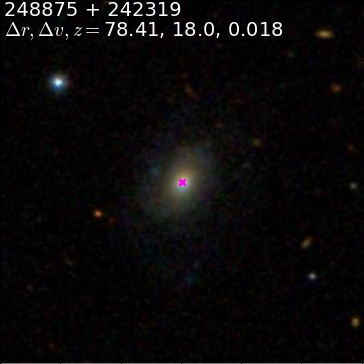}
     \end{subfigure}
     \hfill
     \begin{subfigure}[b]{0.24\textwidth}

         \includegraphics[width=\textwidth]{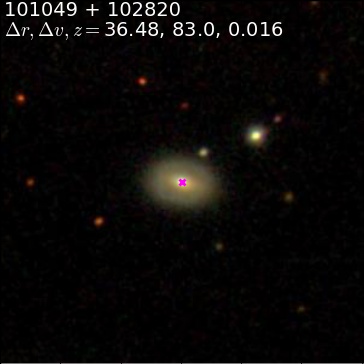}
     \end{subfigure}
     \hfill
     \begin{subfigure}[b]{0.24\textwidth}

         \includegraphics[width=\textwidth]{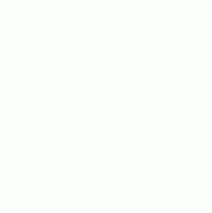}
     \end{subfigure}
     \hfill

\end{figure*}




\bsp	
\label{lastpage}
\end{document}